\documentclass{article}

\usepackage[utf8]{inputenc}
\usepackage{jheppubedit} 

\usepackage{bbold,graphicx,hyperref}
\usepackage{amsfonts,amsmath,amssymb,amsthm,mathtools}                   
\usepackage{color,xcolor}           
\usepackage{float}         
\usepackage{tikz}        
\usepackage{multirow}        
\usepackage{wrapfig}
\usepackage{braket}        
\usepackage{caption}
\usetikzlibrary{shadings,intersections}
\usetikzlibrary{calc,fadings,decorations.pathreplacing}

\newcommand \mathtikz[1] {\quad \vcenter{\hbox{\tikz{#1}}} \quad}

\newcommand\pairA[2]{ 
\begin{scope}[xshift=#1,yshift=#2]
\draw (-0.75,0) -- (-0.25,0) to [out=-90,in=180] (0,-0.33) to [in=-90,out=0] (0.25,0) -- (0.75,0) to [out=-90,in=0] (0,-0.83) to [out=180,in=-90] (-0.75,0);
\end{scope}
}

\newcommand\copairA[2]{ 
\begin{scope}[xshift=#1,yshift=#2]
\draw (-0.75,0) -- (-0.25,0) to [out=90,in=180] (0,0.33) to [in=90,out=0] (0.25,0) -- (0.75,0) to [out=90,in=0] (0,0.83) to [out=180,in=90] (-0.75,0);
\end{scope}
}

\newcommand{\pd}{\partial}
\newcommand{\bea}{\begin{eqnarray}}
\newcommand{\eea}{\end{eqnarray}}
\newcommand{\nn}{\nonumber\\}

\DeclareMathOperator{\tr}{Tr}

\newcommand{\mt}[1]{\textrm{\tiny #1}}

\newcommand{\SL}{\mathrm{SL}}
\newcommand{\PSL}{\mathrm{PSL}}

\title{\boldmath A note on the bulk interpretation of the quantum extremal surface formula}

\author{Gabriel Wong}
\affiliation{Harvard Center of Mathematical Sciences and Applications, USA}
\emailAdd{gabrielwon@gmail.com}

\date{\today}
\abstract{Defining quantum information quantities directly in bulk quantum gravity is a difficult problem due to the fluctuations of spacetime.  Some progress was made recently in \cite{Mertens:2022ujr}, which provided a bulk interpretation of the Bekenstein Hawking formula for two sided BTZ black holes in terms of the entanglement entropy of gravitational edge modes.  We generalize those results  to give a bulk entanglement entropy interpretation of the quantum extremal surface formula in AdS3 gravity, as applied to a single interval in the boundary theory.   Our computation further supports the proposal that AdS3 gravity can be viewed as a topological phase in which the bulk gravity edge modes are anyons that transform under the quantum group $\SL^{+}_{q}(2,\mathbb{R})$.   These edge modes appear when we cut open the Euclidean path integral along bulk co-dimension 2 slices, and satisfies a shrinkable boundary condition which ensures that the Gibbons-Hawking calculation gives the correct state counting.
 }

\begin{document}

\maketitle

\section{Introduction}
Recent progress in the AdS/CFT correspondence has centered on the application of the quantum extremal surface  (QES) formula \cite{Ryu:2006bv,Hubeny:2007xt,Engelhardt:2013tra}. This is a prescription for computing the large N expansion of the entanglement entropy of a boundary subregion $A$  in term of the generalized entropy of  a semi-classical bulk spacetime: 
\begin{align} \label{QES} 
   S_{A} =  S_{\text{gen}} = \frac{\langle{\text{Area}(\gamma_{A})}\rangle }{4G}+ S_{\text{bulk}} +\cdots,
\end{align} 
where $\gamma_{A}$ is the codimension 2 surface that is homologous to the region $A$, defined by extremizing $S_{\text{gen}}$.    The leading $O(N^{2})$  area term is a generalization of the Bekenstein Hawking entropy for black holes. Meanwhile, the subleading $O(1)$  term $S_{\text{bulk}}$ computes the entanglement entropy of quantum fields  across $\gamma_{A}$ in a fixed gravitational background.   The QES formula is the primary tool that has been used to understand how the bulk spacetime geometry  is encoded in the  entanglement structure of the boundary theory.  However its bulk interpretation remains mysterious because the leading area term has no straightforward state counting interpretation.   The main purpose of this article is to provide a bulk canonical interpretation of the QES formula in the setting of AdS3 gravity.  In particular, we will find that the area term can be viewed as counting bulk gravity edge modes.
\paragraph{Bulk modular invariance and the shrinkable boundary condition}
To understand the issue at hand, recall that the generalized entropy is defined via the Euclidean gravity path integral $Z(\beta)$ which sums over spacetimes with a circle $S^{1}_{\infty}$ of length $\beta$ at infinity.    It is defined via the semi classical evaluation of
\begin{align}\label{S}
S&= (1-\beta \pd_{\beta}) \log Z(\beta) \nn
Z(\beta) &\sim  e^{-I_{\text{classical}}} Z_{\text{fluctuations}}
\end{align} 
If $S^{1}_{\infty}$  could be interpreted as a temperature circle, then this is just the standard formula for the thermal entropy of the partition function $Z(\beta)$. However, to obtain the area term in \eqref{QES}
one must evaluate \eqref{S} on a saddle geometry in which $S^{1}_{\infty}$ contracts in the bulk, giving the topology of a cigar \footnote{For example, in the case of the black hole where there is a $U(1)$ symmetry along $S^{1}_{\infty}$, a saddle where the $S^{1}_{\infty}$ remains non contractible in the bulk would have a linear dependence on $\beta$, since the action is independent of the coordinate along the circle.  This would give a linear dependence of $\log Z(\beta)$  on $\beta$, which gives a zero contribution to S. }.   Ironically, the same feature of the geometry which gives the correct value of the entropy also prevents us from giving a straightforward statistical interpretation of \eqref{S}.  

The Euclidean formula \cite{Gibbons:1976ue}  is quite versatile: it is independent of (and predates) the AdS/CFT correspondence,  works for a variety of spinning and charged black holes in various dimensions, and reproduces the Page curve for the Hawking radiation of an evaporating black hole \cite{Almheiri:2019qdq,Penington:2019kki}.   Given that the Euclidean path integral is defined in the low energy effective theory, its success seems somewhat miraculous.    Why does the Euclidean path integral give the correct microstate counting?

\begin{figure}
    \centering
    \includegraphics[scale=.3]{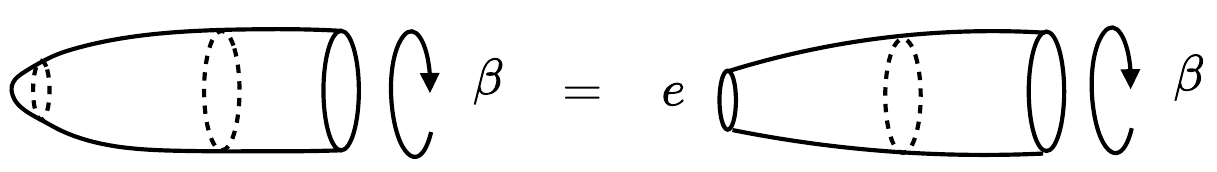}
    \caption{A path integral on geometries with the topology of a cigar can be given a trace interpretation by replacing the tip with a shrinkable boundary condition e.  This effectively replaces the cigar with an annulus, which defines a trace in the ``open string " channel.}
    \label{fig:cigar}
\end{figure}

In \cite{Mertens:2022ujr,Jafferis:2019wkd}, it was argued that the Euclidean calculation gives the right answer because it uses a bulk gravitational version of modular invariance (see also \cite{Mathur:2014nja} for earlier renditions of this argument).    Consider the gravity path integral on geometries with the cigar topology of the Hawking saddle. To obtain a \emph{bulk} trace interpretation, we must remove an infinitesmal disk from the tip of the cigar, and replace it with a ``shrinkable" boundary condition  at the resulting boundary.  This boundary condition is defined so that the path integral on the annulus after the excision equals the path integral on the cigar: see figure \ref{fig:cigar}. 

\begin{figure}[h]
    \centering
    \includegraphics[scale=.3]{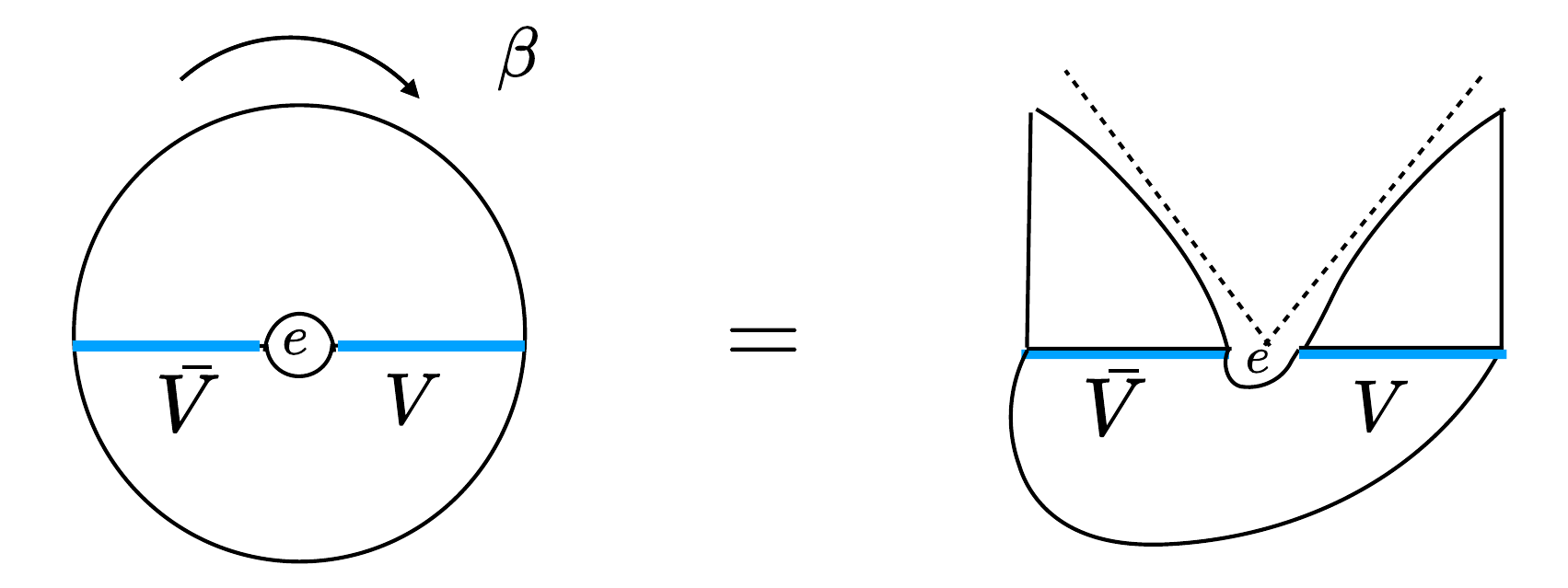}
    \caption{The lower half of the annulus with a shrinkable boundary condition $e$ inserted can be viewed as the path integral preparation of a factorized state in the bulk extended Hilbert space $\mathcal{H}_{V}\otimes \mathcal{H}_{\bar{V}}$.    The subsequent Lorentzian evolution describes an entangled sum of one sided geometries}
    \label{fig:HHsplit}
\end{figure}

Read in the ``open string" channel, the path integral on the annulus seems to define the trace 
\begin{align}\label{Ztr}
Z(\beta)= \tr \rho_{V}^{\beta/2\pi} 
\end{align} 
of a reduced density matrix $\rho_{V}$:  this is shown in figure \ref{fig:HHsplit}.  Given such a reduced density matrix, inserting \eqref{Ztr} back into \eqref{S} shows that the generalized entropy  \eqref{S} is the 
bulk entanglement entropy 
\begin{align}\label{Vn}
S_{gen}= -\tr_{V} \rho_{V} \log \rho_{V} 
\end{align} 
between the subregion $V$ and its complement.  Our main result is an explicit realization of this formula in AdS3 gravity. 

Due to the shrinking cycle, evaluating the bulk path integral  in the open string channel requires the introduction of bulk \emph{edge modes} near the tip of the cigar:  they arise from a local version of holography in which the degrees of freedom of the small disk is replaced by the \emph{gravitational} edge modes at its boundary, where the metric is free to fluctuate in contrast with the asymptotic boundary.    These edge modes are a manifestation of UV-IR mixing in the bulk:  they represent UV degrees of freedom that are normally confined in the IR, but are exposed when we cut open the path integral along a co-dimension 2 entangling surface.   As we will see, their entanglement gives the leading contribution to \eqref{Vn},  which is precisely the area term in the QES formula.


\paragraph{Summary of previous results and outline} 
In this paper we will carry out the argument described above  within the effective theory of AdS3 gravity defined in \cite{Mertens:2022ujr}.  In fact, for the special case of the generalized entropy of the two sided BTZ black hole (with $A$ being the entire spatial slice of a single sided CFT),  equation \eqref{Vn} was derived  in \cite{Mertens:2022ujr}.     In this derivation, a fundamental role was  played by a bulk edge mode symmetry that is intimately related to the modular invariance of the gravity path integral.   Since this same edge mode symmetry governs the QES formula for the single interval EE of a boundary subregion, we summarize those results here.

To begin with, note that in AdS3, the  bulk modular invariance described above is exactly the holographic dual of modular invariance in the boundary CFT.    Here the bulk semiclassical limit corresponds to the high temperature limit when $\beta$ is small.  In this regime we can use the  modular invariance of the CFT to evaluate the partition function in the dual channel: 
\begin{align}
Z(\beta)=Z( \frac{4\pi^{2}}{\beta}) 
\end{align} 
At small $\beta$ the RHS is captured by low energy part of the spectrum, and one usually just keep the vacuum contribution $Z( \frac{4\pi^{2}}{\beta}) \sim \exp \frac{ \pi c}{6\beta} $ - this is an expression of UV-IR mixing in the boundary theory.   The dual bulk calculation corresponds to evaluating the gravity action on the BTZ saddle, where the Euclidean time circle shrinks in the interior as described above.  However, to understand what states are being counted, one should go back to the original channel and expand in the Boltzmann factor $e^{-\beta E}$.   In the standard CFT calculation, this is achieved by doing a Laplace transform, which gives the Cardy density of states $\rho(E)= \exp \sqrt{\frac{c E}{6}}$.   When combined with Brown-Henneaux relation  \cite{Brown:1986nw}  $c=\frac{3 \ell_{AdS}}{2G_{N}}$ this famously reproduces the BH entropy as 
\begin{align}\label{BHC}
\frac{\text{Area}}{4G} = \log \rho (E^*)
\end{align}   
where $E^*$ is the saddle point value of the energy.   In this calculation, BH entropy is naturally interpreted as counting edge modes at asymptotic infinity that transform under the asymptotic Virasoro symmetry.

As explained above, to access the original channel in the bulk, we must introduce gravitational edge modes.  One of the main results of \cite{Mertens:2022ujr} is that these bulk edge modes do not transform under a Virasoro symmetry.  Instead, the shrinkable boundary condition implies that they transform under the quantum group $\SL^{+}_{q}(2,\mathbb{R}) \otimes \SL^{+}_{q}(2,\mathbb{R}) $, with $q$ related to the cosmological constant.  Moreover its co-product naturally defines a factorization of the bulk Hilbert space that leads  to the reduced density matrix $\rho_{V}$ in \eqref{Vn}.   In the classical limit,  equation \eqref{BHC} is then replaced by
\begin{align} \label{dqp2}
\frac{\text{Area}}{4G} = \log \dim_{q} p^*  \dim_{q} \bar{p}^*
\end{align} 
where $p,\bar{p}$ labels representations of $\SL^{+}_{q}(2,\mathbb{R}) \otimes \SL^{+}_{q}(2,\mathbb{R}) $, and $ \dim_{q} p ,  \dim_{q} \bar{p}$ are the associated quantum dimensions which count bulk gravity edge modes.  These results can be viewed as the dimensional uplift of those found in JT gravity in \cite{Blommaert:2018iqz}.

It was originally suggested in \cite{McGough:2013gka} and further advocated in \cite{Mertens:2022ujr} that we should interpret this quantum group symmetry via the paradigm of topological phases, which are long range entangled phases described by topological quantum field theories.  In this context, 
quantum groups arise as the symmetry of anyons that describe the collective degrees of freedom of topological phase.   These anyons are objects in a modular tensor category which defines the TQFT path integral: for Chern Simons theory with compact gauge group $G$,  this modular tensor category is the representation category of a the loop group LG or its associated quantum group.    The results of  \cite{Mertens:2022ujr}  for AdS3 gravity suggests that we should interpret the bulk edge modes as gravitational anyons that belong to the representation category of $\SL^{+}_{q}(2,\mathbb{R}) \otimes \SL^{+}_{q}(2,\mathbb{R}) $, which behaves like a ``gauge group" for gravity.  We will say more about this viewpoint in the conclusion.

Here is a detailed outline of the paper.    
In section \ref{sec:review} we review some salient details of \cite{Mertens:2022ujr} that will be needed for this work.    In \ref{sec:bpf} we define the boundary partition function $Z(\beta)$ and compute its thermal entropy.  In \ref{sec:bgt}, we give the bulk definition of  $Z(\beta)$ in terms of a modified Chern Simons path integral with $\PSL(2,\mathbb{R}) \otimes\PSL(2,\mathbb{R})  $ gauge group \cite{Cotler:2018zff}.   In \ref{sec:bhs} we define the bulk 2 sided Hilbert space for the BTZ black hole geometries and in \ref{sec:btzfact}  we define  $\SL^{+}_{q}(2,\mathbb{R}) $  and  explain how its co-product determines  the factorization of the bulk Hilbert space.   In section \ref{sec:QES}, we repeat these steps for the effective BCFT partition function whose thermal entropy gives the entanglement entropy of a single interval.   We denote  this partition function by $Z_{ee}(l)$, where  $l$ is a dimensionless  length of an interval in the boundary theory, and plays the role of $\beta$ for the black holes.   We derive this partition function carefully in \ref{sec:bdry} by factorizing the boundary theory, and comment on the relation of our entanglement boundary conditions to FZZT branes.  In \ref{sec:BCFTbulk} we give a bulk description of  $Z_{ee}(l)$ using AdS/BCFT and cut this path integral along a Cauchy slice to define a bulk Hartle Hawing state.     In \ref{sec:bulkfact}, we give the factorization of this quantum state and show that its entanglement entropy reproduces the generalized entropy.   In particular, \eqref{dqp2} is now replaced by 
\begin{align}\label{c3l}
\frac{\text{Area}}{4G} = \log \dim_{q} p^*= \frac{c}{3} \log l  ,
\end{align} which gives the entanglement entropy of an interval of length $l$ \footnote{Equation \eqref{c3l} was also obtained in \cite{Lin:2021veu} via a boundary calculation of entanglement entropy in which the Cardy density of states was replaced by $\dim_{q}p$ in an ad hoc fashion.  See section  \ref{sec:bdry} for further details.}.  Here the reduction to a single chiral sector relative to the two sided black hole case can be attributed to introduction of a conformal boundary condition at the entangling surface in the boundary CFT. Finally, in section \ref{sec:ETQFT}, we provide an extended discussion that places our work within the paradigm of extended TQFT.


\section{Review: gravitational edge modes in 3d gravity}
\label{sec:review}
    In this section we review the construction of the effective 3d gravity theory proposed in \cite{Mertens:2022ujr} and its associated gravitational edge modes.     We will see how representations of SL$^+_{q}(2,\mathbb{R})$ naturally arise as the subregion states needed to define the factorization of the bulk Hilbert space of the two-sided BTZ black hole.      
\subsection{The boundary partition function}
\label{sec:bpf}
 Consider an irrational ($c>>1$), holographic 1+1 d CFT with only Virasoro symmetry: the modular invariant partition function is a sum of Virasoro characters:
\begin{align} \label{Zmicro}
    Z(\tau,\bar{\tau})_{\text{micro}}&= \sum_{h,\bar{h}} A_{h,\bar{h}} \chi_{h}(\tau) \chi_{\bar{h}}(\bar{\tau}) ,\qquad \tau =\frac{i \beta}{\ell_{AdS}} (\mu+i) ,
    \end{align}
    where $\beta$ ,$\mu$ is the inverse temperature and chemical potential resp.  
    The Virasoro characters are given by:
\begin{align}
    \chi_0(\tau) = \frac{(1-q)}{\eta(\tau)}q^{-\frac{c-1}{24}}, \qquad \chi_h(\tau) = \frac{1}{\eta(\tau)}q^{h-\frac{c-1}{24}}, \qquad \eta(\tau) \equiv q^{1/24}\prod_{m=1}^{+\infty}(1-q^m),
\end{align}

It is well known \cite{Hartman:2014oaa,Dyer:2016pou, Ghosh:2015iwa} that for theories with gap and a sparse spectrum, the high temperature limit of $Z(\tau,\bar{\tau})$ is dominated by the vacuum character 
in the dual channel
\begin{align}
    Z(\tau,\bar{\tau})_{\text{micro}}  =\sum_{h,\bar{h}} A_{h,\bar{h}} \chi_{h}(-1/\tau) \chi_{\bar{h}}(-1/\bar{\tau})  \sim |\chi_{0}(-1/\tau)|^2
\end{align}
Specifically, the vacuum block dominates when the inverse temperature is much smaller than the gap to the first excited state:
\begin{align}
\frac{\beta}{\ell_{AdS}} << \Delta_{\text{gap}}
\end{align}

This motivates us to define an effective theory with partition function:
\begin{align}\label{Ztau}
    Z(\tau,\bar{\tau}):= |\chi_{0}(-1/\tau)|^2
\end{align}
This partition function can be viewed as an effective theory which describes the universal dynamics of an irrational, Virasoro CFT in the high temperature limit.  

To see the statistical interpretation of $Z(\tau,\bar{\tau})$, we go back to the original channel-where the temperature circle is small- by applying a modular transformation to express the vacuum character $\chi_{0}(-1/\tau)$ as a sum over non-degenerate Virasoro characters $\chi_{h}(\tau)$. It will be convenient to use the Liouville parameterization\footnote{It should be emphasized that our boundary theory is NOT Liouville theory, whose partition function is a diagonal sum of all non degenerate Virasoro characters} of the conformal dimensions,
\begin{align}
    h &= p^2 + Q^2/4,\quad \bar{h} = \bar{p}^2 + Q^2/4, \quad
    Q=b +b^{-1}, \quad c= 1+6 Q^{2}.
\end{align} Then we can use the Virasoro S-matrix \cite{Zamolodchikov:1995aa} to write 
\begin{align}\label{mod}
    Z(\tau,\bar{\tau})=|\chi_{0}(-1/\tau)|^2 &= \int_0^{+\infty} \int_0^{+\infty} \hspace{-0.2cm} dp \,\,d \bar{p} \quad 
    S_{0}{}^{p}S_{0}{}^{\bar{p}} \chi_{p}(\tau) \chi_{\bar{p}}(\bar{\tau}),
\end{align}
where
\begin{equation}
S_{ 0}{}^{p} = \sqrt{2} \sinh ( 2 \pi b p ) \sinh ( 2 \pi b^{-1} p)
\end{equation}

As noted in \cite{McGough:2013gka} and further explained below, these Virasoro S-matrix elements form a measure on the space of Virasoro representations, which is identical to the Plancherel measure for SL$_{q}^{+}(2,\mathbb{R})$
\begin{align}
    S_{ 0}{}^{p}= \dim_{q}p, \qquad q =e^{i \pi b^{2}}
\end{align}
The equality of these measures is the first hint that Rep (SL$_{q}^{+}(2,\mathbb{R}$) )might play a role in 3d gravity.

In terms of the inverse temperature $\beta$ and chemical potential $\mu$ defined in \eqref{Zmicro}, we can write $ Z(\tau,\bar{\tau})$ explicitly as the partition function 
\begin{align}
\label{solidtorus}
&Z(\beta,\mu) \equiv \text{Tr}\left[ e^{-\beta H +i\mu \frac{\beta}{\ell} J}\right]  \\
&=\int_0^{+\infty} \int_0^{+\infty} \hspace{-0.2cm} dp \,\,d \bar{p}  \dim_{q}(p) \dim_{q} (\bar{p})  \frac{e^{-\frac{\beta}{\ell}(p^2 + \bar{p}^2)} e^{i\mu \frac{\beta}{\ell}(p^2 - \bar{p}^2)}}{\left|\eta(\tau)\right|^2}, \nonumber
\end{align}
The integral runs over primary states labelled by $(p,\bar{p})$, with primary energy and angular momentum given by 
\begin{align}\label{MJ}
H &= \frac{p^2 + \bar{p}^2}{\ell} ,\qquad
J = p^2-\bar{p}^2
\end{align}   As we explain below, these can be identified with black hole states in the bulk, whereas the descendants states captured by the Dedekind eta are the boundary gravitons.

The thermal entropy of the boundary partition function \eqref{solidtorus}  reproduces the BH entropy for Euclidean BTZ black holes in the high temperature $\beta/l_{\text{AdS}} <<1$, and  semi-classical limit $c>>1 \to b>>1$ limit.   For $\beta/l_{\text{AdS}} <<1$, large $p,\bar{p} $ dominates the integral so we can approximate:
\begin{align}
    \dim_{q} p \sim \exp 2 \pi bp = \exp  \sqrt{\frac{cL_{0}}{6}}
\end{align}
This gives the expected Cardy density of states.  For $c>>1$, we can evaluate the entropy at the saddle point  $(p^*,\bar{p}^*)$, which gives 
\begin{align}
   S&=(1-\beta \pd_{\beta}) \log Z(\beta,\mu)\nn
   &\sim \log \dim_{q}p^* \dim_{q}\bar{p}^*= \frac{A(M^*,J^*)}{4G} 
   \end{align}  
  Here $(M^*,J^*)= (\frac{p^{*2} + \bar{p}^{*2}}{\ell} ,p^{*2}-\bar{p}^{*2})  $ are the saddle point values of the mass and spin, which depends on $\beta$.   

\subsection{The bulk gravity theory} 
\label{sec:bgt}
 The boundary partition function \eqref{solidtorus} has a dual interpretation in terms of 3d pure gravity with a negative cosmological constant. The gravity action is:
\begin{align}
I_{\text{EH}} = \frac{1}{16\pi G_{\mt{N}}^{(3)}} \int d^3x \sqrt{-g}(R^{(3)}- \frac{1}{l^{2}_{\text{AdS}}}) ,
\end{align}
where the bulk and boundary parameters are related via the Brown Hennaux relation $c= \frac{3l_{\text{AdS}}}{2G_{N}}$. 

This duality can be derived explicitly in the first order formulation of 
AdS3 gravity as a Chern Simons theory with  gauge group $\PSL(2,\mathbb{R}) \otimes \PSL(2,\mathbb{R})$, in which the gauge fields are related to the Vielbein $e$  and spin connection $\omega$ according to 
\begin{align}\label{AA}
    A= \frac{e}{\ell} + \omega ,\, \quad  \bar{A} =\frac{e}{\ell}- \omega
\end{align}  Specifically, \cite{Cotler:2018zff} showed that the Euclidean Chern Simons path integral with gravitational boundary condition on a solid torus with a contractible Euclidean time cycle reproduces the vacuum character $|\chi_{0}|^{2}(-1/ \tau) $. Using coordinates $\varphi,t_{E}$ on the boundary and $\rho$ as the bulk radial coordinate, the gravitational boundary conditions are:\footnote{ To describe the boundary theory in terms of group elements $g$, one also has to impose a $\PSL(2,\mathbb{R})$ quotient $g \sim h(\rho) g$ since these define the same gauge field $A=g^{-1} dg $ } 
\begin{enumerate}
    \item Asymptotic AdS3 boundary conditions that set the boundary value of the gauge field to 
    \begin{align} \label{adsbc}
    A_{\varphi}=A_{t_{E}}&= \begin{pmatrix} 0 && \mathcal{L} (\varphi,t_{E})\\ 1&&0 \end{pmatrix} \nn
    A_{t_{E}} - A_{\varphi}&=0,
    \end{align} where $\mathcal{L} (\varphi+it_{E})$ is a classical boundary stress tensor which generates the boundary Virasoro algebra with $L_{n}= \int \frac{d \varphi}{2 \pi } e^{i n (\varphi)} \mathcal{L} (\varphi )$.
    \item A winding constraint defined as follows.    Writing the flat connection on the $(\rho, T_{E})$ disk as $A=g^{-1} dg$, one requires that at the asymptotic  boundary , the map 
    \begin{align}
    g \big|_{\rho=\infty}: S^{1} \to  \PSL(2,\mathbb{R})
    \end{align}  winds once around the non contractible $SO(2)$ subgroup of $\PSL(2,\mathbb{R})$. 
    \end{enumerate}

\begin{figure}[h]
    \centering
    \includegraphics[scale=.3]{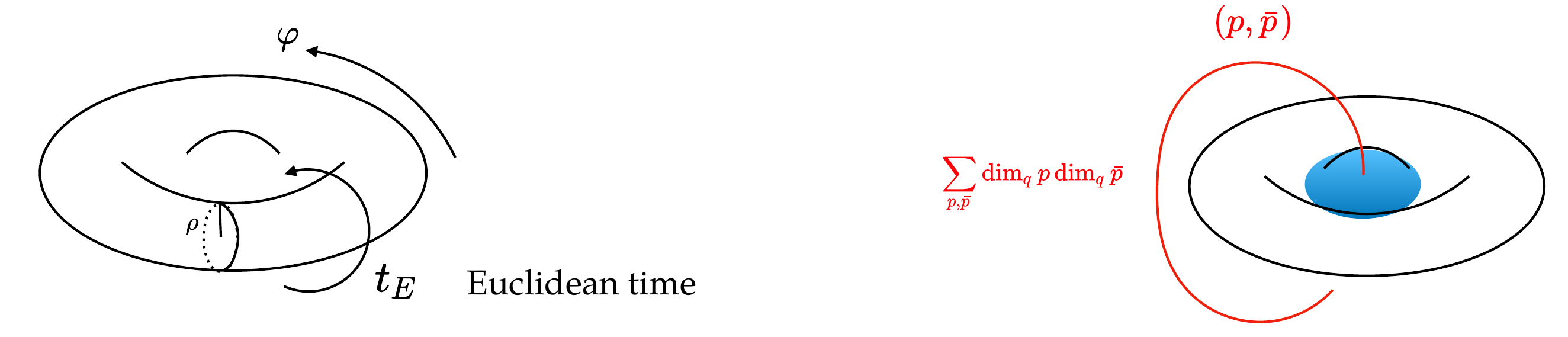}
    \caption{On the left, we have the solid torus partition function of the bulk TQFT, with no Wilson lines inserted.  This can be given a perturbative interpretation as the quantization of fluctuations around a BTZ saddle obtained by filling the Euclidean time cycle.  On the right, we interpret $Z(\tau,\bar{\tau}) $  in terms of the solid torus in the dual channel, where we fill in the spatial cycle with a disk  punctured by a superposition of Wilson lines.   The punctured disk is the spatial slice of a black hole geometry describing individual microstates of the thermal ensemble \eqref{solidtorus}}
    \label{fig:solidT}
\end{figure}

In the standard description \cite{Maloney:2007ud,Cotler:2018zff}, the bulk gravity path integral is given a perturbative interpretation as arising from the quantization of fluctuations about a Euclidean BTZ saddle, where the Euclidean time circle contracts smoothly in the bulk: this is shown in the left of figure \ref{fig:solidT}, which is a solid torus with no Wilson loops inserted.  This perspective is made explicit by writing
\begin{align}\label{pert}
    Z(\tau,\bar{\tau}) = |\chi_{0}(-1/\tau)|^2 = \big|\exp(\frac{-2 \pi i }{\tau} \frac{c}{24}\big)\prod_{n=2}^{\infty}\frac{1}{1- \exp( \frac{-2 \pi i n}{\tau}) }|^{2},
\end{align}
where $\exp(\frac{2 \pi i }{\tau} \frac{c}{24}\big)$ is the dominant contribution from the on shell Euclidean classical action, and $\prod_{n=2}^{\infty}|\frac{1}{1- \exp( \frac{2 \pi i n}{\tau}) }|^{2}$ is the partition function for fluctuations associated with boundary gravitons.  In this description, the BH entropy is captured geometrically by the horizon area of the saddle point geometry, but it does not have a state counting interpretation.   This is because the bulk time slice is an annulus $\mathcal{A}$, and the solid torus does not take the form $\mathcal{A}\times S^1$.   

The bulk generalized entropy is computed by including the contributions of the one loop determinant about the BTZ saddle to the BH entropy; in this case the generalized entropy gives the exact thermal entropy of  $Z(\tau,\bar{\tau})$, because the partition function is one loop exact \cite{Maloney:2007ud}:
\begin{align}\label{Sgen}
    S_{\text{gen}} = (1-\beta \pd_{\beta}) \log Z(\beta,\mu)
\end{align}
While the perturbative description  \eqref{pert}of $Z(\tau,\bar{\tau})$ is correct, it has the undesirable feature that only a small part of the total entropy -  the part due to fluctuations around the BTZ saddle- has a state counting interpretation.   The heavy primaries in \eqref{solidtorus} which gives the dominant contribution to the entropy is hidden in this bulk description.  
One way to incorporate these heavy primaries into a \emph{bulk} statistical mechanical partition function is to go to the dual channel via a modular transform, as we did in \eqref{mod}.  In this channel,  the bulk is obtained by filling in the \emph{spatial} $\varphi$ cycle with a disk, which evolves in the Euclidean time direction to give a bulk trace (see right of figure \ref{fig:solidT}).  The modular transform introduces Wilson lines labelled by the primaries $(p,\bar{p})$, which punctures the spatial  disk and creates a nontrivial holonomy around $\varphi$, leading to the nondegenerate characters $\chi_{p}(\tau)$.  The bulk is thus a sum over solid torus path integral with Wilson loops inserted, and leads to a \emph{non-perturbative}\footnote{By non-perturbative, we mean not expanding around a Euclidean saddle.  We do not mean that we are including terms non perturbative in the gravitational coupling} description of $Z(\tau,\bar{\tau})$ as a statistical sum \eqref{solidtorus} .   For the real values of $p,\bar{p}$ which appear in \eqref{solidtorus}, the classical phase space on the punctured disk\footnote{ In description of the bulk Hilbert space in terms of Liouville conformal blocks, these punctures correspond to the insertion of a Liouville vertex operator with conformal dimension $h=Q^{2}/4 +p^{2}$ } consists of flat connections with holonomies in the hyperbolic conjugacy class.  In the metric description, these hyperbolic holonomies correspond to \emph{Lorentzian}  BTZ black holes with mass an spin given by \eqref{MJ}.   Quantizing these solutions\footnote{The exact Hilbert space has black holes with quantized angular momenta, which links $p$ and $\bar{p}$ so the two sectors do not decouple. However our effective description at high temperature can not resolve this discreteness, so we get two independent chiral and anti chiral sectors } gives rise to the black hole states which is counted by the partition function \eqref{solidtorus}.  These black holes states should not be confused with the Euclidean saddle: the former are ``microstates" with zero entropy and no temperature, while the latter is geometric representation of the saddle point of \eqref{solidtorus} with mass and spin determined by the temperature $\beta$.

\begin{figure}[h]
    \centering
    \includegraphics[width=0.75\textwidth]{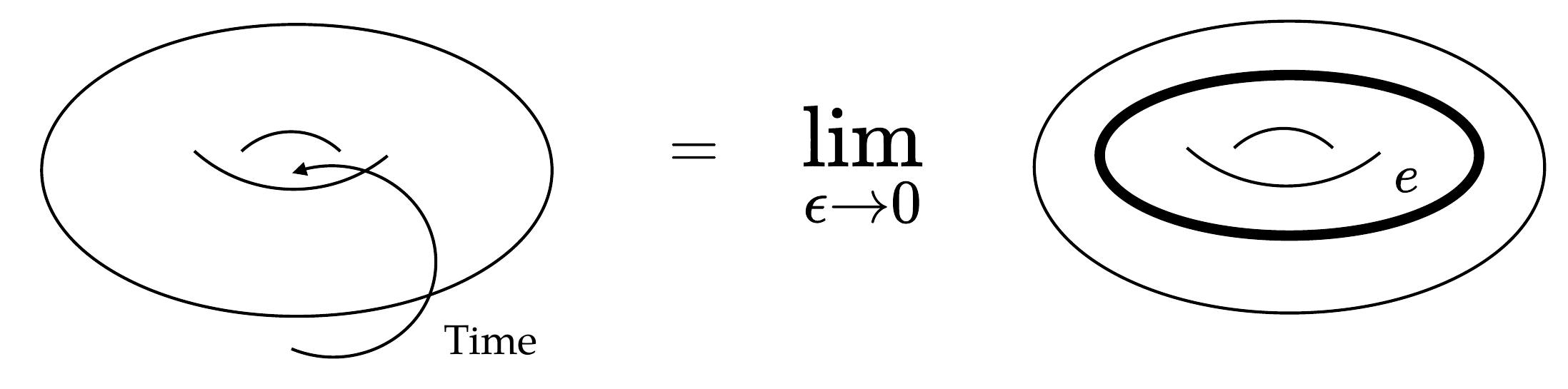}
    \caption{The shrinkable boundary condition}
    \label{fig:dfacthh}
\end{figure}

Thus, the dual channel gives a perfectly consistent definition of a bulk statistical partition function.  However to obtain an entanglement entropy interpretation of $S_{gen}$, we need to give a trace interpretation in the channel on the left of figure \ref{fig:solidT}, where we filled in the time cycle with no Wilson loops inserted. To do this, we need to remove a thin tube from the center of the of the solid torus, with a shrinkable boundary condition $e$ (figure \ref{fig:dfacthh}).  The tube arises from the modular evolution of a stretched entangling surface: in Lorentzian signature, this would correspond to a stretched horizon as depicted in the right of figure \ref{fig:HHsplit}.     The shrinkable boundary condition ensures that the resulting path integral can be given a state counting interpretation as
\begin{align} \label{shrink}
Z(\tau,\bar{\tau})= \text{tr}_{V} \rho_{V}, 
\end{align} where $\rho_{V}$ is a \emph{bulk} reduced density matrix on a subregion $V$.   Note that since the dynamical spacetime geometry is encoded into the field space via \eqref{AA}, we can define the subregion $V$ in a topological sense, without fixing the metric at the entangling surface.   This is crucial for obtaining the correct gravity edge modes, which differ from the conventional gauge theory edge modes that do fix a background metric at the entangling surface.  Finally, we observe that the shrinkable boundary condition $\eqref{shrink}$ guarantees that the bulk generalized entropy, identified with the full thermal entropy of $Z(\tau,\bar{\tau})$ via \eqref{Sgen}, is given by the entanglement entropy of $\rho_{V}$:
\begin{align}
    S_{\text{gen}} = -\tr_{V} \rho_{V} \log \rho_{V}
\end{align}
In the next section, we review the bulk Hilbert space and factorization map which gives rise to the reduced density matrix $\rho_{V}$. 

Before moving on to the discussion of bulk factorization, we should emphasize that the gravity boundary conditions- in particular the winding constraint- alter Chern Simons gauge theory in an essential way. One is no longer using the path integral measure associated to the gauge group $\PSL(2,\mathbb{R}) \otimes \PSL(2,\mathbb{R})$, which would naturally sum over integer winding numbers.   In fact, the main proposal of  \cite{Mertens:2022ujr} is that the bulk theory should be viewed as an extended TQFT associated with the quantum semi group $\SL_{q}(2,\mathbb{R})^+ \otimes \SL_{q}(2,\mathbb{R})^+ $- we will comment on this perspective in the section \ref{sec:ETQFT}.   To avoid confusion, we will henceforth refer to the bulk theory as the gravitational TQFT, or just the bulk TQFT for short.

\subsection{The bulk Hilbert space}
\label{sec:bhs}
\begin{figure}
    \centering
    \includegraphics[scale=.5]{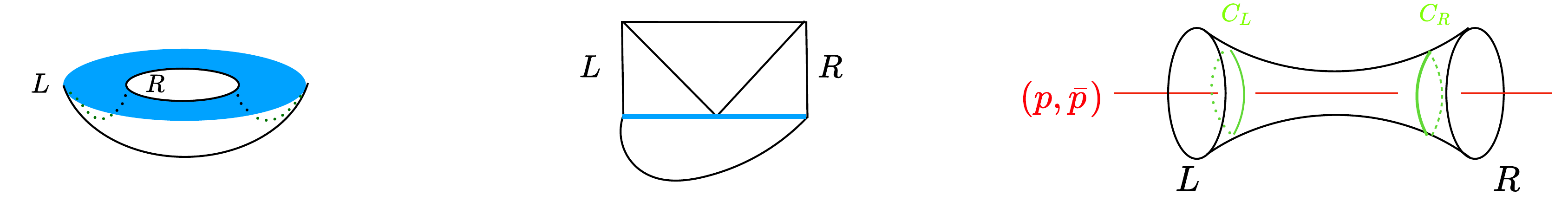}
    \caption{On the left we show the Einsten Rosen bridge as the time refection symmetric slice of the Euclidean BTZ geometry. On the right, we show the two sided Lorentzian geometry of obtained by evolving the Hartle Hawking state produced by the path integral  on the left}
    \label{fig:annulus}
\end{figure}

We start by defining the bulk Hilbert space on the time-reflection symmetric Cauchy slice of the solid torus geometry (with no Wilson loops inserted).  This is an annulus (depicted in figure \ref{fig:annulus})
 which corresponds to the Einsten Rosen bridge in the Lorentzian continuation of the geometry.
The Hilbert space is obtained by quantizing the  classical phase space solutions with the same topology as the two sided BTZ black holes.   These correspond to the the Kruskal extension of the single-sided Banados geometries:
   \begin{align}
    ds^{2}&= l^{2}_{\text{AdS}} \frac{dr^{2}}{r^2}- (r dx^+ -  l^{2}_{\text{AdS}}\frac{\bar{\mathcal{L}}(x^-) d x^-}{r})(r dx^- -  l^{2}_{\text{AdS}}\frac{\mathcal{L}(x^+) d x^+}{r}),\nn
    x^{\pm}&= t \pm \varphi
\end{align}
which are parameterized by the functions $\mathcal{L} (x^+), \bar{\mathcal{L}}(x^-)$. These are the  the expectation values of the boundary stress tensors which determine the boundary values of the gauge fields $A_{\varphi} ,\bar{A}_{\varphi}$ according to equation \eqref{adsbc}.

The Kruskal extension introduces a left (L) and right (R) asymptotic boundary, with coordinates $x^{\pm}_{L},x^{\pm}_{R} $ respectively. This gives 4 stress tensor components:
\begin{align} \label{stress}
    \mathcal{L}(x^+_{L}),\,\, \bar{\mathcal{L}}(x^-_{L}), \,\, \mathcal{L} (x^+_{R}), \,\,\bar{\mathcal{L}}(x^-_{R})
\end{align} 
The L and R copies are not quite independent, because they share the same zero mode which determines the mass and spin of the black hole.  This correlation between the L and R boundaries is most easily seen in the Chern Simons representation, where the zero modes label the holonomy around the non-contractible spatial cycle according to:
\begin{align} 
2\cosh (\frac{p}{2}) = \tr P \exp \oint_{C} A_{\varphi} d\varphi\nn
2\cosh (\frac{\bar{p}}{2}) = \tr P \exp \oint_{C} \bar{A}_{\varphi} d\varphi
\end{align}
   Heuristically, we can think of this holonomy as measuring the effect of Wilson lines that thread the wormhole in the dual channel (as in left of figure \ref{fig:ERsplit} ). The shared zero mode implies that the phase space does not factorize into independent copies associated to the two boundaries \footnote{  In fact, in addition to the the four stress tensors in\eqref{stress}, this is an additional degree of freedom in the gravity phase space parametrized by a radial Wilson line connecting the two boundaries \cite{Henneaux:2019sjx}.  This is needed to obtain a non-degenerate symplectic form in the bulk.}

Due to the AdS3 boundary conditions \eqref{adsbc},  the holonomy labels $(p,\bar{p})$ can be identified with the primary labels for the Virasoro algebra.
Upon quantization of the stress tensor degrees of freedom  in \eqref{stress} , we obtain a bulk Hilbert space $\mathcal{H}_{\text{bulk}}$ constructed out of Virasoro primaries:
\begin{align} \label{Hbulk}
\mathcal{H}_{\text{bulk}} 
&=\text{span} \{\ket{p, m^{L}_{k},m_{k}^{R}} \otimes  \ket{\bar{p}, \bar{n}_{k}^{L},\bar{n}_{k}^{R}}\},
\end{align} 
 where $p,\bar{p}$ labels the black hole mass and spin according to \eqref{MJ} , while
 $m^{L}_{k},m_{k}^{R}$ and $\bar{n}_{k}^{L},\bar{n}_{k}^{R}$ are the occupation numbers of the descendant states corresponding to boundary gravitons.  Explicit we can write:
 \begin{align}\label{pbasis}
     \ket{p, m^{L}_{k},m_{k}^{R}}= \prod_{k=1}^{\infty} L_{-k}^{m^{L}_{k}} \ket{p}_{L}\otimes \prod_{k=1}^{\infty} L_{-k}^{m^{R}_{k}}\ket{p}_{R}
 \end{align}
 While the L and R edge modes corresponding to the boundary gravitons factorize, the bulk degrees of freedom 
$(p, \bar{p})$ prevent a complete factorization into independent, one-sided Hilbert spaces.  

The Hartle Hawking state, prepared by the half solid torus path integral, can be expressed explicitly as 
\begin{align} \label{HartleH}
\ket{\Psi(\beta,\mu)}&=\ket{HH} \otimes \ket{\bar{HH}}\nn
\ket{HH}&=\int_{0}^{\infty} dp\,\,\sqrt{\dim_{q}p} \sum_{m^{L}_{k}=m^{R}_{k}} e^{ \frac{-\beta}{\ell}(-1+i \mu )(p^{2} + N_{m_{k}})} \ket{p\,m^{L}_{k}\,m^{R}_{k}}
\end{align} 
and satisfies $Z(\beta,\mu)= \braket{\Psi(\beta,\mu)|\Psi(\beta,\mu)}$.
\subsection{Factorization and half-wormhole states}
\label{sec:btzfact}
\begin{figure}[h]
    \centering
    \includegraphics[scale=.3]{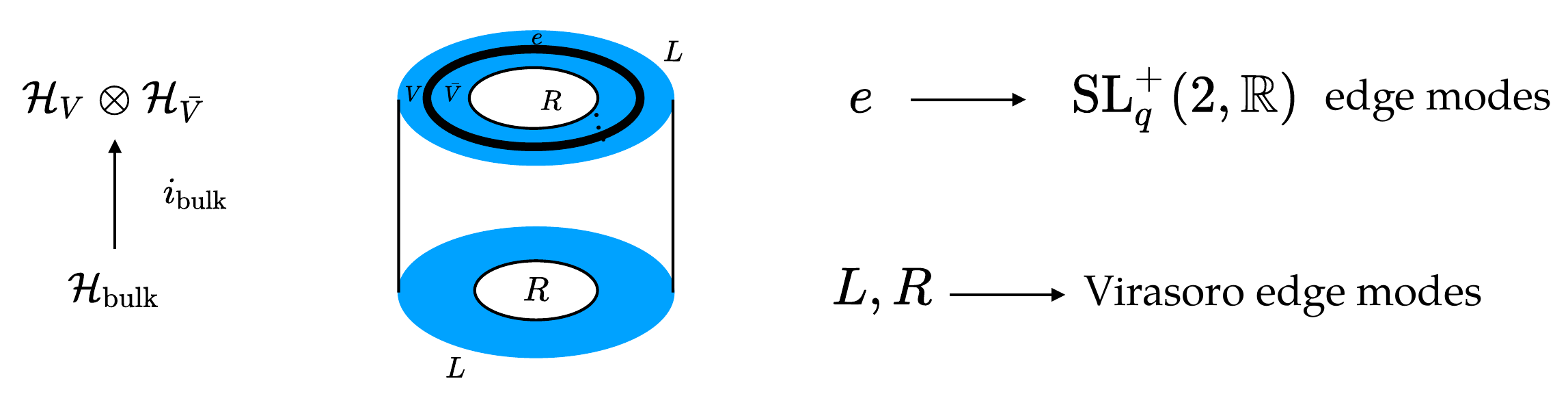}
    \caption{A path integral process that factorizes the ER bridge}
    \label{fig:annfact}
\end{figure}

\begin{figure}[h]
    \centering
    \includegraphics[scale=.3]{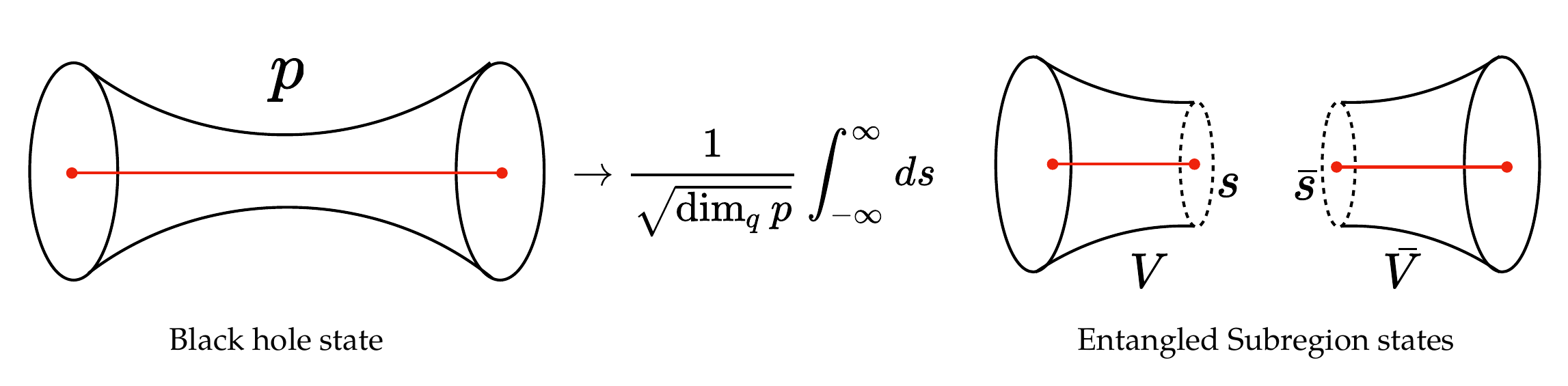}
    \caption{A heuristic depiction of the bulk factorization map via a bulk version of ER=EPR.}
    \label{fig:ERsplit}
\end{figure}

In this section we define the factorization map from the Hilbert space on the ER bridge into a \emph{bulk} extended Hilbert space
\begin{align}
  i_{\text{bulk}}:  \mathcal{H}_{\text{bulk}} \to \mathcal{H}_{V} \otimes \mathcal{H}_{\bar{V}} ,
\end{align} 
where  $\mathcal{H}_{V}$,$\mathcal{H}_{\bar{V}} $, are the subregion Hilbert spaces on each half of the ER bridge as shown in figure \ref{fig:ERsplit} .   $ i_{\text{bulk}}$ can be viewed as a bulk path integral evolution that splits the ER bridge into two half wormholes: this process is depicted in figure \ref{fig:annfact}.
It was explained in \cite{Mertens:2022ujr} that compatibility with the shrinkable boundary condition requires that each of subregions $V$ supports Virasoro edge modes at asymptotic infinity, and quantum group edge modes  at the entangling surface transforming under $\SL_{q}^{+}(2,\mathbb{R}) \otimes \SL^{+}_{q}(2,\mathbb{R})$.

The most direct, though formal way of defining $\SL_{q}^{+}(2,\mathbb{R}) $ is via its fundamental (co) representation as a two by two ``quantum" matrix: 
\begin{align}
   g\equiv \begin{pmatrix}a&&b\\c&&d
    \end{pmatrix}
\end{align}
in which each matrix element $a,b,c,d$ is an operator on $L^{2}(\mathbb{R}\otimes \mathbb{R})$ with \emph{positive spectrum}.   These operators have commutation relations given by
\begin{align} \label{msl2commutators}
ab&= q^{1/2} ba, \quad  ac= q^{1/2} ca,\quad  bd= q^{1/2} db,\quad cd= q^{1/2} dc,  \nn
bc&=cb,\quad ad-da = (q^{1/2}-q^{-1/2}) bc.
\end{align} 
When $q\to 1$, these matrix elements can be taken to be real numbers, and the positive spectrum condition corresponds to requiring that they are positive real numbers.  This limit gives the semi group SL$^{+}(2,\mathbb{R})$ : due to positivity, not all elements have inverses \footnote{The $q \to 1$ limit of our 3d gravity theory gives JT gravity, which was formulated as a ``SL$^{+}(2,\mathbb{R})$ gauge theory" in  \cite{Blommaert:2018iqz}.   There it was explained that the SL$^{+}(2,\mathbb{R})$ measure removes gauge field configurations that would correspond to singular metrics that are not included in the gravity path integral.  The q-deformed SL$_{q}^{+}(2,\mathbb{R})$ measure plays a similar role in removing unwanted metric configurations from the measure: see section  5.3 of \cite{Mertens:2022ujr} for more details  }.  Likewise, when $q \neq 1$, the positive spectrum condition implies that the quantum  matrix $g$ does not always have an antipode, which is the analogue of an inverse for a quantum group.  In this sense SL$_{q}^{+}(2,\mathbb{R})$ is a \emph{quantum semi-group}.  

This definition may seem a bit bizarre at first sight: in what sense do the elements $g$ define a symmetry?     The simplest answer to this question is that a symmetry G should be defined by its representation category Rep(G), and to say that SL$_{q}^{+}(2,\mathbb{R})$ is a symmetry means there  exists a well defined category of SL$_{q}^{+}(2,\mathbb{R})$ representations, related to the continuous series representations of SL$_{q}(2,\mathbb{R})$.  In fact the (highly nontrivial) representation matrices $R_{ab}(g) \,\, a, b \in \mathbb{R} $ for SL$_{q}^{+}(2,\mathbb{R})$ have been worked out explicitly by \cite{ip2012representation}.  The indices $a,b$ are continuous but the matrix elements satisfy the usual representation property, which says that if $g=g_{1}g_{2}$, then
 \begin{align}\label{rep1}
      R_{ab}(g) = R_{ab}(g_{1}\cdot g_{2}) = \int dc \,\,R_{ac}(g_{1})R_{cb}(g_{2})
 \end{align}

One of the main results of \cite{Mertens:2022ujr} is that the factorization map $i_{\text{bulk}}$ can be characterized algebraically as the co-product of the quantum group $\SL_{q}^{+}(2,\mathbb{R}) \otimes \SL_{q}(2,\mathbb{R})^{+}$. This co product is equivalent to the  representation property \eqref{rep1}, viewed as a map from functions of one quantum group element $g$ to functions on two quantum group elements $g_{1},g_{2}$.   The matrix element $R_{ab}(g)$ is wavefunction of the ``Wilson line" variable $g$ that crosses the ER bridge, $g_{1} ,g_{2}$ are the subregion Wilson lines obtained by the factorization map.  The quantum group edge modes correspond to the contracted indices $c$.  Below we give a condensed explanation  of how these representation matrices and their co-product arise in the factorization of  $\mathcal{H}_{\text{bulk}}$.  We refer the reader to \cite{Mertens:2022ujr} for further details.  
\paragraph{Factorization} 

To set up the discussion, it is useful to organize our bulk Hilbert space in terms of Virasoro representations $V_{p}$ as follows.  First, the Hilbert space decouples into a tensor product chiral and anti chiral parts, corresponding to the two $\PSL(2,\mathbb{R})$ components of the bulk gauge field:
\begin{align}
    \mathcal{H}_{\text{bulk}} &= \mathcal{H} \otimes \bar{\mathcal{H}}
\end{align}
We will focus just on a single sector $\mathcal{H}$, which takes the form 
 \begin{align}\label{H}
  \mathcal{H}= \oplus_{p}V_{p} \otimes V_{p}^*
 \end{align} 
 with basis given in \eqref{pbasis}.  It turns out $i_{\text{bulk}}$ acts only on the bulk zero mode subspace $\mathcal{H}^{0} \subset \mathcal{H}$ , i.e. the primary states $\ket{p}$ in \eqref{pbasis}. The boundary gravitons labelled by $m_{k}^{L,R}$ are spectators.  Moreover, no descendants are introduced at the bulk entangling surface: in this sense the shrinkable boundary condition is a gapped or topological boundary condition.  This is in contrast with the usual factorization map for Chern Simons theory, which would introduce descendants at the bulk entangling surface.  In the shrinking limit, these states lead to an infinite entanglement entropy  associated with the UV divergence of QFT on a fixed background.  Gravity regularizes this divergence.

The crucial observation that underlies our factorization map is the existence of 
a one to one correspondence between the Virasoro representations $V_{p}$ and representations $\tilde{V}_{p}$ of the quantum group $\SL_{q}^{+}(2,\mathbb{R}) $.
This is a non compact generalization of the well known relation\footnote{In mathematics this is called the Kazhdan-Lusztig equivalence \cite{Kazhdan1993TensorSA}} between representations of the loop group LG and corresponding  quantum group, usually denoted $\mathcal{U}_{q}(G)$. \footnote{ $\mathcal{U}_{q}(G)$ refers to the q-deformation of the universal enveloping algebra of $G$, which is just the algebra generated by the Lie algebra elements subjected to the commutation relations.   }   
The correspondence between $V_{p}$ and $\tilde{V}_{p}$ is a deep fact linked to the modular bootstrap for irrational CFT's with $c>1$. The modular bootstrap is a set of constraints that determines the data for the representation category Rep(Vir) of the Virasoro algebra: These include the spectrum of primaries, the fusion rules, and the OPE coefficients.   Ponsot and Teschner \cite{Ponsot:1999uf} showed that there is a one to one map $\mathcal{F}$  ( a ``functor") between representation categories
\begin{align}\label{functor}
\mathcal{F}:    \text{Rep (Vir)} \to \text{ Rep(SL}_{q}^{+}(2,\mathbb{R}))
\end{align}
 such that the fusion algebra of Rep (Vir ) maps to  representation ring of $\SL_{q}^{+}(2,\mathbb{R})$. In particular, this fusion rules of Rep(Vir) corresponds to the decomposition of tensor products in  Rep($\SL_{q}^{+}(2,\mathbb{R})$), and the Virasoro Fusion matrix maps to the Racah Coefficents (6j symbols) of    Rep($\SL_{q}^{+}(2,\mathbb{R})$).   Thus, Rep($\SL_{q}^{+}(2,\mathbb{R})$) provides a solution to the modular bootstrap. 

Starting from this observation, we can define the factorization map $i_{\text{bulk}}$ acting on the zero mode subspace  $\mathcal{H}^{0}$ as follows

\begin{enumerate}
    \item First we use the correspondence \eqref{functor} to relate primaries in $\mathcal{H}^{0}$ to representation matrix elements of $SL_{q}^{+}(2,\mathbb{R})$. Concretely, we introduce frozen quantum group indices $\mathfrak{i}_{L},\mathfrak{i}_{R}$ to the primary states:
    \begin{align}
        \ket{p}\to \ket{p ,\mathfrak{i}_{L},\mathfrak{i}_{R}} \in \mathcal{H}^{0}
    \end{align}
    such that the corresponding zero mode wavefunctions on the ER bridge are given by representation matrix element of $\SL_{q}^{+}(2,\mathbb{R})$
    \begin{align}\label{qbasis}
      \braket{ g| p ,\mathfrak{i}_{L},\mathfrak{i}_{R}} = \sqrt{\dim_{q} p} \,\,R^{p}_{\mathfrak{i}_{L},\mathfrak{i}_{R}}(g) ,\qquad g\in \SL_{q}^{+}(2,\mathbb{R})
    \end{align} 
  The normalization is important: it corresponds to the Plancherel measure on $SL_{q}^{+}(2,\mathbb{R})$, which we explain below. 
 The frozen indices originate from the PSL$(2,\mathbb{R})$ Kacs-Moody zero modes of the WZW model which arises from the usual bulk-boundary correspondence for   Chern Simons theory.  However,  the AdS3 boundary conditions \eqref{adsbc} projects onto a particular linear combination of Kacs-Moody zero modes, labelled by $\mathfrak{i}_{L,R}$, thus reducing the symmetry algebra to Virasoro.  The projection restricts to a subspace of wavefunctions invariant (up to a phase) under the action of right (left) multiplication by elements  $h_{R/L} \in H_{R/L}$ of certain subgroups of $ \SL_{q}^{+}(2,\mathbb{R})$:
\begin{align}
     R^{p}_{ a \mathfrak{i}_{R}}(g )\to R^{p}_{a\mathfrak{i}_{R}}(g h_{R})&= e^{i \gamma(h_{R}))} R^{p}_{a\mathfrak{i}_{R}}(g) \qquad h_{R} \in H_{R}\nn
      R^{p}_{  \mathfrak{i}_{L}b}(g )\to R^{p}_{\mathfrak{i}_{L}b}( h_{L}g)&= e^{i \gamma(h_{L}))} R^{p}_{ \mathfrak{i}_{L}b}(g)\qquad h_{L} \in H_{L}
\end{align} 
The projected matrix elements are called Whittaker functions, and the associated quantum state only depends on the configuration space variable $g\in H_{L} \backslash \SL_{q}^{+}(2,\mathbb{R})/H_{R}$ in the double coset
    \item We can give a more precise characterization of $\mathcal{H}^{0}$.  This follows from the fact that the matrix elements
    \begin{align}
    R^{p}_{ab}(g),\qquad  a,b, \in \mathbb{R},\quad p \in \mathbb{R}^+
    \end{align} form a complete basis on the space  $\text{L}^{2}(\SL^+_{q}(2,\mathbb{R})) $ of integrable functions on $\SL_{q}^{+}(2,\mathbb{R})$.  This completeness is due to a nontrivial generalization of the Peter Weyl theorem \cite{Ponsot:1999uf} that states that  $\text{L}^{2}(\SL^+_{q}(2,\mathbb{R})) $ decomposes into representations $\tilde{V}_{p}\otimes \tilde{V}_{p}^*$ of $\SL^+_{q}(2,\mathbb{R})$ (the $^*$ denotes the dual vector space):
    \begin{align} \label{L2}
\text{L}^{2}(\SL^+_{q}(2,\mathbb{R})) &= \int_{\oplus_{p \geq 0} }   \,d \mu(p)\, \tilde{V}_{p} \otimes \tilde{V}^{*}_{p}\nn
 d\mu(p) &= \dim_{q}p  \qquad q= e^{\pi i b^{2}}.
    \end{align}
The bulk zero mode Hilbert space is obtained from \eqref{L2} by projecting the representation matrix elements onto invariant subspaces labelled by $\mathfrak{i}_{L,R}$. This is expressed by a modified Peter Weyl theorem: 
\begin{align}\label{L20}
\mathcal{H}^{0}&= L^{2} ( H_{L}\backslash \SL^+_{q}(2,\mathbb{R}) / H_{R} ) = \int_{\oplus_{p \geq 0} }   \,d \mu(p)\, \tilde{V}_{p,\mathfrak{i}_{L}} \otimes \tilde{V}^{*}_{p,\mathfrak{i}_{R}},
\end{align} 
where  $\tilde{V}_{p,\mathfrak{i}_{L}}$,$\tilde{V}^{*}_{p,\mathfrak{i}_{R}}$ denote the respective invariant subspaces. 
Note the similarity with the structure of \eqref{H}: the correlation in the representation label $p$ gives rise to the $L^{2}$ space, allowing us to interpret the zero mode subspace $\mathcal{H}^{0}$  as wavefunctions on the quantum group:  in this sense $g \in \SL^+_{q}(2,\mathbb{R})$ is a ``Wilson line" variable that captures the connectedness of the ER bridge. 

Now some technical remarks about the measure in \eqref{L2} and \eqref{L20}. 
The Plancherel measure $d\mu(p) = \dim_{q}(p)$ is defined by the norm of the matrix elements with respect to the Haar measure, and appears in the orthogonality relation: 
 \begin{align} 
 \int_{\SL^+_{q}(2,\mathbb{R})} d g\,   R^{p}_{ab}(g)  R^{p*}_{cd}(g) =  \frac{1}{\dim_{q} p}\, \delta_{bd} \delta_{ac}\,.
 \end{align} 
 This explains the normalization in \eqref{qbasis}.  The appearance of the Plancherel measure in \eqref{L2} means that the completeness relation is given by
 \begin{align} 
 \delta(g-g') = \int d\mu (p)\,\int da \int db  \, R^{p}_{ab}(g)R^{p*}_{ab}(g').
 \end{align} 
 Moreover, it is identified with the character in $\tilde{V}_{p}$ evaluated on the identity element: 
 \begin{align} \label{count}
 \tr_{\tilde{V}_{p}} (\bold{1}) = \dim_{q} p 
 \end{align} 
Note that the Plancherel measure is a highly constrained part of the Rep($\SL^+_{q}(2,\mathbb{R})) $data.  In particular, an arbitrary rescaling of the representation matrices would spoil the representation property
 \begin{align}\label{rep}
      R^{p}_{ab}(g_{1}\cdot g_{2}) = \int dc \,\,R^{p}_{ac}(g_{1})R^{p}_{cb}(g_{2})
 \end{align}
 \item In addition to a (non-commutative) pointwise product $\text{L}^{2}(\SL^+_{q}(2,\mathbb{R})) $ also has a co-product obtained by pulling back the multiplication rule on $\SL^+_{q}(2,\mathbb{R})$ : 
 \begin{align}
  \Delta: \text{L}^{2}(\SL^+_{q}(2,\mathbb{R}))  &\to  \text{L}^{2}(\SL^+_{q}(2,\mathbb{R}))  \otimes \text{L}^{2}(\SL^+_{q}(2,\mathbb{R}))  \nn
   f(g) &\to f(g_{1}\cdot g_{2}) 
 \end{align}
    When applied to the representation basis , this defines a factorization map that ``cuts open" the Wilson line via the representation property \eqref{rep}.
  \item  Finally,  we use the co product $\Delta$ to define the factorization map on our bulk Hilbert space.  The action on the zero modes is:
  \begin{align}\label{fact0}
  i_{\text{bulk}}: \mathcal{H}^{0} &\to \mathcal{H}^{0}_{V}\otimes \mathcal{H}^{0}_{\bar{V}}\nn
 R^{p}_{\mathfrak{i}_{L},\mathfrak{i}_{R}}(g) &\to R^{p}_{\mathfrak{i}_{L},\mathfrak{i}_{R}}(g_{V}g_{\bar{V}}) =\int ds R^{p}_{\mathfrak{i}_{L} s }(g_{V}) R^{p}_{\mathfrak{i}_{L} s} (g_{\bar{V}})\nn
 \ket{p \,\mathfrak{i}_{L} \,\mathfrak{i}_{R}} &\to \frac{1}{\sqrt{\dim_{q}p}} \int_{-\infty}^{\infty} \,ds \,\ket{p \,\mathfrak{i}_{L} \, s}\otimes \ket{p \,s \,\mathfrak{i}_{R}}
 \end{align}

 Here we have defined the subregion zero mode subspaces  spanned by the respective matrix elements as 
     \begin{align}
      \mathcal{H}^{0}_{V} &=  L^{2} ( H_{L} \backslash \SL^+_{q}(2,\mathbb{R})  ) =\int_{\oplus p\in \mathbb{R}^+}  d \mu(p)  \tilde{V}^{*}_{p,\mathfrak{i}_{L}} \otimes  \tilde{V}_{p}\nn
      \mathcal{H}^{0}_{\bar{V}} &= L^{2} ( \SL^+_{q}(2,\mathbb{R})/H_{R}  ) = \int_{\oplus p\in \mathbb{R}^+}  d \mu(p)  \tilde{V}_{p}\otimes \tilde{V}^{*}_{p,\mathfrak{i}_{R}}
     \end{align} 
 The group elements  $g,g_{V},g_{\bar{V}}$ in \eqref{fact0}belong to the respective cosets above.   We can lift the factorization map onto the full Hilbert space $\mathcal{H}$ by tensoring $ \tilde{V}^{*}_{p,\mathfrak{i}_{L}} $ with the  descendant states, which just gives back the Virasoro representation $V_{p}$ . Thus we can express the spectral decomposition of the subregion Hilbert spaces in a simple way:
 \begin{align}\label{sub}
 \mathcal{H}_{V} &= \int_{\oplus p\in \mathbb{R}^+}  d \mu(p) V_{p} \otimes  \tilde{V}_{p} = \text{span} \{ \ket{p \, \mathfrak{i}_{L}\, s : m_{k}^{L}} \} \nn
  \mathcal{H}_{\bar{V}} & \int_{\oplus p\in \mathbb{R}^+}  d \mu(p)  \tilde{V}_{p} \otimes  V_{p}= \text{span} \{ \ket{p \, \mathfrak{i}_{R}\, s : m_{k}^{R}} \} 
 \end{align}

 The factorization map is then given by 
 \begin{align}\label{ibulk}
   i_{\text{bulk}}: \mathcal{H}&\to \mathcal{H}_{V}\otimes \mathcal{H}_{\bar{V}}\nn
  \ket{p \,\mathfrak{i}_{L} \,\mathfrak{i}_{R}:m_{k}^{L} \, m_{k}^{R}} &\to \frac{1}{\sqrt{\dim_{q}p}} \int_{-\infty}^{\infty} \,ds \,\ket{p \,\mathfrak{i}_{L} \, s:m_{k}^{L}}\otimes \ket{p \,s \,\mathfrak{i}_{R}:m_{k}^{R}}
 \end{align} 
 \end{enumerate} 
 
\paragraph{Summary}
In this section, we explained aspects of the representation theory of  $\SL^+_{q}(2,\mathbb{R})$ and defined a bulk  factorization map via the co product on $\SL^+_{q}(2,\mathbb{R})$ .   While the discussion above may have seemed formal, it has a very simple and intuitive pay off.     Heuristically, we can view the configuration space variable  $g$  as a $\SL^+_{q}(2,\mathbb{R})$  Wilson line  $g=P\exp \int A$ that crosses the ER bridge.    The factorization map $i_{\text{bulk}}$ is then defined from splitting this Wilson line  $g \to g_{V}g_{\bar{V}}$ into subregion Wilson lines in a manner analogous to the splitting of an ordinary Wilson line according to \eqref{rep} ( see figure \ref{fig:ERsplit}). 

Finally, let us comment on the semi classical description of  our Hartle Hawking state, which is a particular superposition of these quantum group Wilson lines.
In \eqref{HartleH}, we gave an abstract description of this state in terms of entangled sums of Virasoro primaries and descendants. 
 However, there is a more geometric description that was recently analyzed in \cite{Chua:2023ios},\cite{Chua:2023srl}.
To apply their analysis to our Hartle Hawking states \eqref{HartleH}, we first use \eqref{MJ} to relabel the basis states  
\begin{align} 
\ket{p,\mathfrak{i}_{L},\mathfrak{i}_{R}} \ket{\bar{p},\mathfrak{i}_{L},\mathfrak{i}_{R}} \to \ket{E,J},
\end{align}
where E the energy and J the angular momentum of a micro-canonical black hole state.  Generalizing the JT story,  the authors  \cite{Chua:2023ios},\cite{Chua:2023srl} showed that the overlap 
\begin{align}
\braket{E,J | \Psi(\beta,\mu)} 
\end{align} 
can be computing semiclassically via a mixed boundary value problem in which certain components of the spatial metric and conjugate variables to the remaining components are fixed.  With these boundary conditions, they evaluated the on shell action $I[E,J]$ for these configurations and found
\begin{align}
    \braket{E,J | \Psi(\beta,\mu)} \simeq e^{-I} = e^{S(E,J)/2 - \beta E/2} ,
\end{align}
with $S$ the black hole entropy.
This is exactly the semi classical limit of our Hartle Hawking wavefunction when projected on associated  primary state (The entropy $S(E,J)$ coming from the Plancherel measures $\sqrt {\dim_{q}p}$ ).

\section{The single interval QES formula from bulk edge modes} 
\label{sec:QES}
In this section we use the gravitational TQFT to give a bulk derivation of the QES formula for the vacuum state of a boundary theory on $S^{1}\times \mathbb{R}$ . At first sight, it may seem strange to apply our bulk TQFT, designed  to capture the high temperature behavior of the boundary theory, to a zero temperature state. However, the computation of entanglement entropy for a subregion accesses the high temperature regime because of the Unruh effect. Observers confined to a spacetime subregion of the boundary theory must accelerate. As a result they see a thermal state, and the associated thermal entropy is dominated by the high temperature region close to the entangling surface.   The entanglement entropy of the sub-region is equal to the associated thermal entropy.

To show the parallel structure between the bulk derivation of the RT formula and the BH entropy, we will follow the same sequence of steps as in section  \ref{sec:review}.  We start with the properly factorized boundary calculation, and write down the effective partition function whose thermal entropy gives the entanglement entropy of an interval. We then give the bulk interpretation and define the bulk factorization map.   Finally, we show how the semi-classical limit of the bulk entanglement entropy reproduces the RT formula.
\subsection{The boundary calculation of entanglement entropy for one interval}
\label{sec:bdry}
\begin{figure}
    \centering
    \includegraphics[width=0.75\textwidth]{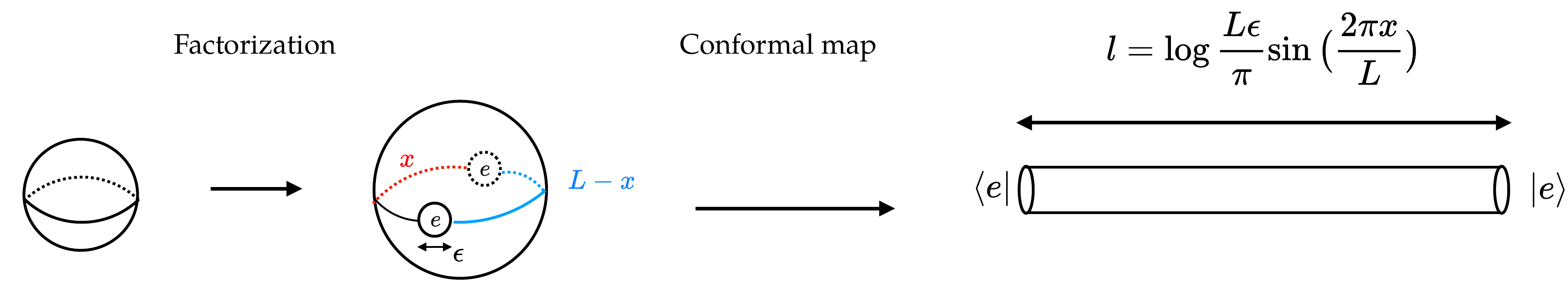}
    \caption{From the point of view of the effective partition function for the subregion observer, the factorization map introduces two holes in the sphere partition function, where the entanglement boundary state $\ket{e}$ is inserted.  }
    \label{fig:CFTcyl}
\end{figure}

We begin by reviewing the canonical calculation of entanglement entropy for an interval $A$ in the vacuum state $\ket{0}\in \mathcal{H}_{S^1}$ of a CFT on a spatial circle \cite{Hung:2019bnq} of length $L$.    The vacuum is prepared by the disk which ends on this circle, and its norm is the sphere partition function.   To properly factorize the vacuum state, we introduce a regulator surface around the entangling surface, on which we choose a shrinkable boundary condition \cite{Hung:2019bnq}  \cite{Donnelly:2018ppr}. The factorize state is then prepared by a path integral on a hemisphere with two semi-disks removed. This can be read as an ``open string" amplitude between the intervals $A$ and its complement $\bar{A}$, with respective lengths $x$ and $L-x$. \begin{figure}
    \centering
    \includegraphics[width=0.75\textwidth]{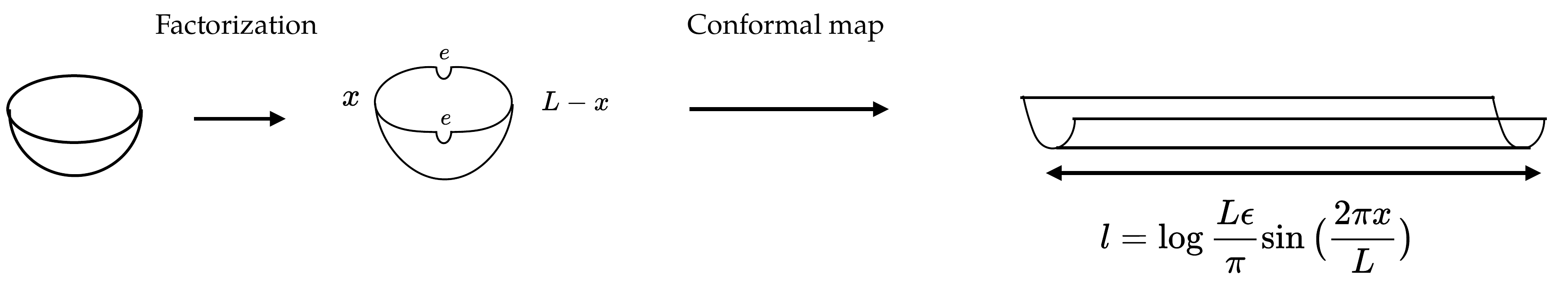}
    \caption{The factorization of the CFT vacuum introduces regulator surfaces near the endpoints of the interval.  After a conformal map, this gives a half annulus}
    \label{fig:CFTHH}
\end{figure}
This is shown in figure  \ref{fig:CFTHH}, where we have made a conformal map to a long strip of length 
\begin{align}
    l = \log \frac{L \epsilon}{\pi}  \sin \big( \frac{2 \pi x}{L}) \big)  
\end{align}
To understand the nature of the shrinkable boundary condition, it is useful to consider the trace of the reduced density matrix $\rho_{A}$ on $A$.  As shown in figure \ref{fig:CFTcyl},  $\tr_{A} \rho_{A} $  is identified with an annulus, which we can view as a ``closed string" amplitude between two entanglement boundary states $\ket{e}$.   In the limit $\epsilon \to 0$, this boundary state must be defined so that annulus reproduces the sphere partition function. 

 Strictly speaking, any boundary state with non trivial overlap with the vacuum will satisfy the shrinkability condition in the strict $\epsilon \to 0$ limit.  This is because this limit correspond to the long time propagation that projects to the vacuum, which is the path integral on a small disk that closes the hole.   However, the vacuum is not a conformal invariant boundary state.   In general, a conformally invariant boundary condition is given by a superposition of Ishibashi states $\ket{h}\rangle$ satisfying
\begin{align}
    ( L_{n} - \bar{L}_{-n}) \ket{h}\rangle =0 ,
\end{align}
with $h$ ranging over the spectrum of the CFT on the circle. The requirements of conformal symmetry and shrinkability then implies we should choose a boundary state $\ket{e}\in \mathcal{H}_{S^1}$ with nonzero overlap $\langle \braket{0|e} $.   In a holographic CFT, the assumption of a gap and a sparse spectrum implies that projecting to $\ket{0} \rangle$ gives a good approximation.   This is because for an arbitrary conformally invariant boundary state $\ket{a}$,
\begin{align}
    \braket{a| \tilde{q}^{L_{0} + \bar{L}_{0} -\frac{c}{24}}| a} &=  \sum_{h} |\braket{a|h}|^{2} \langle \braket{h |\tilde{q}^{ L_{0}+\bar{L}_{0} -\frac{c}{24}} |h}\rangle \nn
    &= \sum_{h}  |\braket{a|h}|^{2}  \chi_{h}(\tilde{q}) ,
\end{align}
so as $\tilde{q} \to 0 $, the vacuum character in the dominates.   Following the approach in \cite{Hung:2019bnq}, we choose $\ket{e}=\ket{0}\rangle $ as the entanglement boundary state.   

Inserting this boundary state at the regulated entangling surface defines an effective partition function $Z_{ee}(l)$ given by  
\begin{align}\label{Zee}
  Z_{ee}(l) =\langle \braket{0 |\tilde{q}^{ L_{0}+\bar{L}_{0} -\frac{c+\bar{c}}{24}} |0}\rangle =\chi_{0}(\tilde{q}) ,\qquad \tilde{q}=\exp (-2l).
\end{align}
We have thus arrived once again at the vacuum character, this time with only one chirality.
Note that we have applied the BCFT analogue of the logic used in section 2, which used modular invariance and the presence of a gap to argue for the dominance of the vacuum block in the dual channel.  In the BCFT setting, the dual channel is the closed string channel.
 
In the shrinking limit $l \to \infty$,  the long time propagation projects onto the vacuum, so we can write ( setting $c=\bar{c}$)
\begin{align}\label{cl12}
    Z_{ee}(l)= \exp(\frac{c l}{12})
\end{align}
This is exactly the form of the sphere partition function as dictated by the conformal anomaly, showing explicitly that shrinkability is satisified.   
In this limit, the entanglement entropy 
\begin{align}
    S_{A}=(1-n\pd_{n}) \log Z_{ee}(\frac{l}{n})\to &=\frac{c}{6} l \nn
    &= \frac{c}{6}\log \frac{L \epsilon}{\pi}  \sin \big( \frac{2 \pi x}{L}) \big)
\end{align}
gives the standard universal answer.  Notice that in the closed string channel,  replicating the cylinder actually maps the parameter $l \to \frac{l}{n} $, because the definition of closed string Hamiltonian $L_{0}+\bar{L}_{0}$ requires us to rescale the geometry so that the circumference of the cylinder has length $2\pi$.

\paragraph{State counting in the open string channel} 
To understand what states are being counted by the entanglement entropy $S_{A}$ we need to go back to the open string channel and define the reduced density matrix.   We do so by applying a modular transformation using the Virasoro S matrix
\begin{align}\label{Zl}
    Z_{ee}(l)= \chi_{0}(\tilde{q}) &= \int \, dp\, S_{0}^{p} \,\chi_{p}(q)= \int_{0}^{\infty} dp \,\, \dim_{q}(p)  e^{-\frac{2 \pi^{2}}{l} p^{2}} \frac{1}{\eta(q)}\nn
    q&=e^{\frac{-2 \pi^{2}}{l}}
\end{align}
This equation defines the statistical partition function   $Z(l)$ on an interval of length $l$,with $l$ playing the role of a dimensionless temperature.  
The sum is over Virasoro representations labelled by $p$ with a measure given by $S_{0}^p= \dim_{q}p$.  While non strictly necessary,  it is useful to apply the language of Liouville theory to give a concrete interpretation of the subregion Hilbert space $\mathcal{H}_{A}$ on which the partition sum \eqref{Zl} is defined.  This is  because Liouville theory with brane insertions gives a local field theory description for Virasoro representations.  In particular,  for each $p$, we can view $\chi_{p}(q)$ as the Liouville partition function  on a spatial interval with a ZZ brane inserted at one, and FZZT branes at the other.     This corresponds to a quantization of Liouville theory on the interval with local boundary conditions \cite{Dorn:1994xn,Zamolodchikov:1995aa}.   The different characters $\chi_{p}$ in \eqref{Zl} are obtained by quantizing with different FZZT branes, also labeled by $p$.    We can thus view the integral over $p$ in \eqref{Zl} as an integral over boundary conditions with measure $\dim_{q}p$.

\begin{figure}[h]
    \centering
    \includegraphics[width=0.75\textwidth]{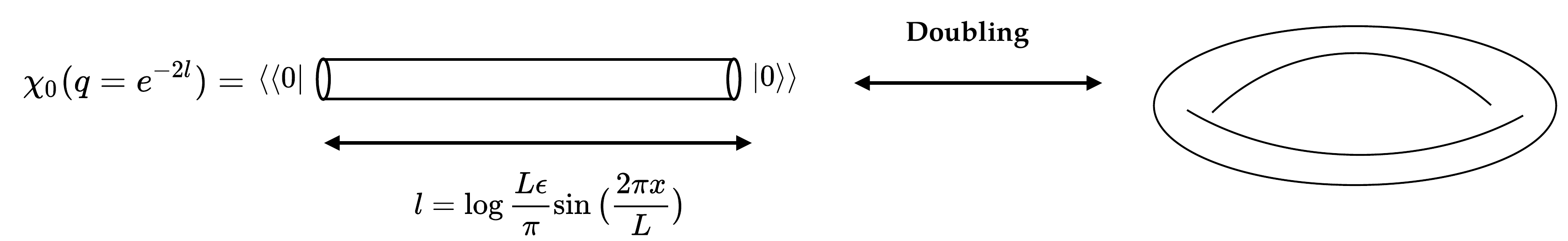}
    \caption{Using the doubling trick replaces the long cylinder with a thin torus.  The partition function $Z_{ee}(l)$ is the chiral vacuum conformal block on the torus. }
    \label{fig:doubling}
\end{figure}

The end result is that Hilbert space $\mathcal{H}_{A}$ on an interval with entanglement boundary conditions can be view as a direct sum of Virasoro representations 
\begin{align}
    \mathcal{H}_{A}= \oplus_{p\geq 0} \,\, V_{p} \qquad  V_{p}=\text{span} \{ \ket{p\,m_{k}} \} ,\qquad  \braket{p\,m_{k} | p' m_{k}'} =\delta (p-p') \delta_{m_{k} m_{k}'} 
\end{align}
where as before $p$ labels primaries and $m_{k}$ the occupation number of the descendants. 
Equivalently, we can apply the unfolding trick\footnote{This is a standard procedure in BCFT.  On the half plane, the conformally invariant boundary condition given by the Ishibashi states can be expressed as \begin{align} \big( T(z)-\bar{T}(\bar{z})\big) \ket{0}\rangle=0\end{align}.  This condition on the stress tensor is exactly what is needed to analytically continue $T(z)$ into the lower half plane such that $T(z^*)= \bar{T}(\bar{z})$.  In the description of the full plane there is only a single chiral stress tensor $T(z)$ and no boundary.   } to the annulus and view each $\chi_{p}(q)$ as a chiral conformal block on a torus (see figure \ref{fig:doubling})- this will be particularly useful when we consider the holographic dual.   
In the unfolded theory, we interpret \eqref{Zl} as a sum over states on a circle, whose $Z_{2}$ quotient reproduces the original interval.  Either way we find that the factorization of the global vacuum state $\ket{0}\in \mathcal{H}_{S^1}$ that is compatibility with shrinkable boundary condition is  
\begin{align}
     &i_{\text{boundary}} :\mathcal{H}_{S^1} \to \mathcal{H}_{A}\otimes \mathcal{H}_{\bar{A}}\nn
   &i_{\text{boundary}}\quad :\ket{0} \to \ket{HH}= \frac{1}{\sqrt{Z_{ee}}}\int_{0}^{\infty} dp\,\,\sqrt{\dim_{q}p} \sum_{m_{k}} e^{ \frac{-\pi^{2}}{l}(p^{2} + N_{m_{k}})} \ket{p,m_{k}}_{A} \otimes \ket{p,m_{k}}_{\bar{A}}
\end{align}
$N_{m_{k}}=\sum k m_{k} $ is the total level for the occupation numbers $m_{k}$. We have denoted the factorized state as $\ket{HH}$, because we will see below that it corresponds to a bulk Hartle Hawking state. The reduced density matrix on $A$ is
\begin{align}
    \rho_{A}= \frac{1}{Z_{ee}(l)} \int_{0}^{\infty} dp\,\,\dim_{q}p \sum_{m_{k}} e^{ \frac{-2\pi^{2}}{l}(p^{2} + N_{m_{k}})} \ket{p,m_{k}}\otimes \bra{p,m_{k}}
\end{align}
and satisfies $Z(l)= \tr_{A} \rho_{A}$ by design.
Notice that $l$ now plays the role that $\beta/\ell_{AdS}$  did in the black hole thermal density matrix.  Since we are taking the limit $l \to \infty$ as $\epsilon \to 0$,  we can apply an analogous high temperature saddle point approximation as was done in the black hole case.   
In particular, in the $l \to \infty$ limit, the $p$- integral  for $Z(l)$ is dominated by large $p$. Further taking a $c>>1$ classical limit, corrsponding to $b>>1$, we can approximate $\dim_{q}p \sim \exp 2 \pi bp$ and find the saddle point 
\begin{align}
    p^*=\frac{bl}{2\pi}
\end{align}
This gives $Z_{ee}(l) \sim  \exp \frac{c l}{6} $
which agrees with the closed string channel calculation \eqref{cl12}.  The fact that a saddle point approximation of the effective partition function $Z_{ee}(l)$ reproduces the RT formula was observed in \cite{Lin:2021veu}. However, in that work, the $\dim_{q}p$ measure was introduced in an ad hoc fashion, based on its agreement with the Cardy density of states at high energies.  Our work provides a justification for the calculation in \cite{Lin:2021veu}.

\subsection{Bulk partition function and Hilbert space} 
\label{sec:BCFTbulk}

Let us now consider the bulk description of the partition function $Z_{ee}(l)$.  If we use the unfolded representation of $Z_{ee}(l)$ as the chiral vacuum block on a boundary torus, we can essentially repeat our analysis of the bulk path integral for the black hole spacetimes in section 3.  In particular, the relevant  bulk dual is obtained by filling in the modular time circle of the torus with a smooth disk. Indeed, the same path integral derivation used in \cite{Cotler:2018zff} shows that the solid torus path integral with a single copy of the gauge field reproduces the vacuum character $\chi_{0}(-1/\tau)$ on the boundary.

\begin{figure}[h]
    \centering
    \includegraphics[width=0.75\textwidth]{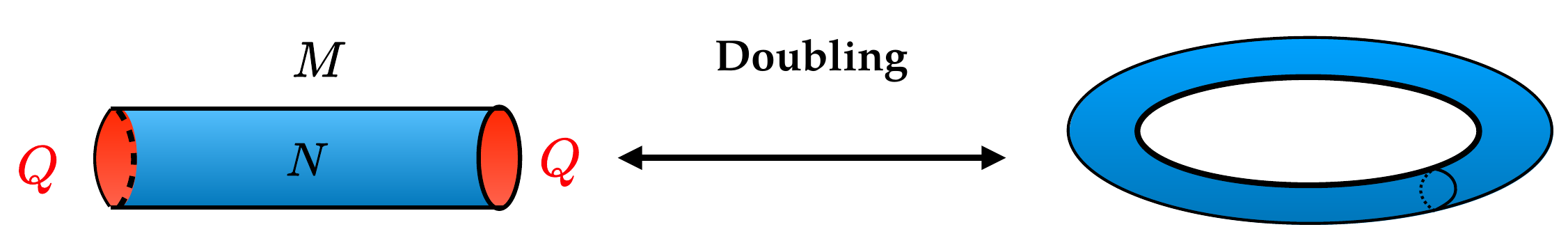}
    \caption{The figure shows the gravitational description of the effective partition function $Z_{ee}(l)$.  On the right we use the doubled representation where we fill in the modular time circle.  On the left, we show the bulk dual in the cylinder representation of the boundary, denoted by $M$ in the figure.  The bulk dual is a manifold $N$ whose boundary is  $\pd N= Q \cup M$.  Q represent end of the world branes in a bulk dual that is locally AdS3.}
    \label{fig:bulkdouble}
\end{figure}

However, in the original boundary cylinder description of $Z_{ee}(l)$, the bulk dual has a more complicated description.  Denoting the cylinder manifold by $M$, a bulk manifold that ends on $M$ must include end-of-the-world-branes \cite{Takayanagi:2011zk}.   An example of such a manifold is shown in the left of figure \ref{fig:bulkdouble}.   Here the bulk manifold is $N \cup Q$, where $N$ is the bulk spacetime and $Q$ the end of the world branes that are caps which extend from the regulator surface $\pd M$  into the bulk.  The bulk spacetime satisfies $\pd N= M \cup Q$.   The end of the world branes Q are responsible for reducing the asymptotic symmetry of the bulk space time to a single copy of Virasoro, consistent with the fact that the conformal boundary conditions on $M$ breaks the Virasoro symmetry to a single chiral copy.

A concrete description of the action for these branes has been given in the literature on AdS/BCFT \cite{Fujita:2011fp} .  In the metric formulation they modify the Einstein Hilbert action\footnote{There is also a standard Gibbons Hawking action at the asymptotic boundary $M$, which we omit below} $I_{EH}$ to
\begin{align}
    I_{EH} &\to I_{EH}+ I_{GH}+I_{brane}\nn
    I_{GH} &=\int_{Q} \sqrt{h} K \nn
    I_{\text{brane} } &=T\int_{Q} \sqrt{h}
\end{align}
where $h_{ab}$ is the induced metric on the brane $Q$, $T$ is the brane tension\footnote{When the boundary CFT satisifes a local boundary condition given by a Cardy state $\ket{a}$, the brane tension $T$ is tuned so that it gives a contribution to the entanglement entropy that  matches with boundary entropy of the boundary CFT.   The boundary entropy of a BCFT is defined to be $g_{a} =\log \braket{a|0}$. }, and $K=h^{ab}K_{ab}$  is the trace of the extrinsic curvarture on $Q$. This is defined in terms of the unit normal $n_{a}$ on $Q$ as $K_{ab} =\nabla_{a}n_{b} $.  The Gibbons Hawking term allows for a variational principle that is consistent with either Neumann or Dirichlet boundary conditions on the metric: since the ETW brane Q lives in the bulk, its metric should be free to fluctuate. Thus, Neumann BC is chosen there, corresponding to
\begin{align}
    K_{ab} + Kh_{ab}= T h_{ab} 
\end{align}

\begin{figure}[h]
    \centering
      \includegraphics[width=0.55\textwidth]{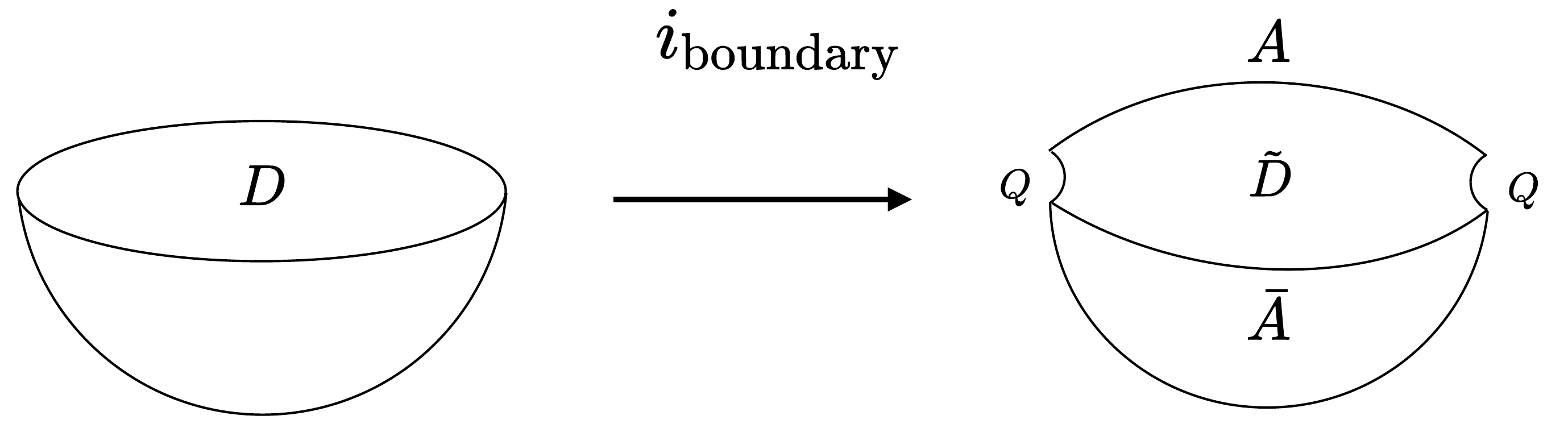}\qquad \qquad  
    \includegraphics[width=0.35\textwidth]{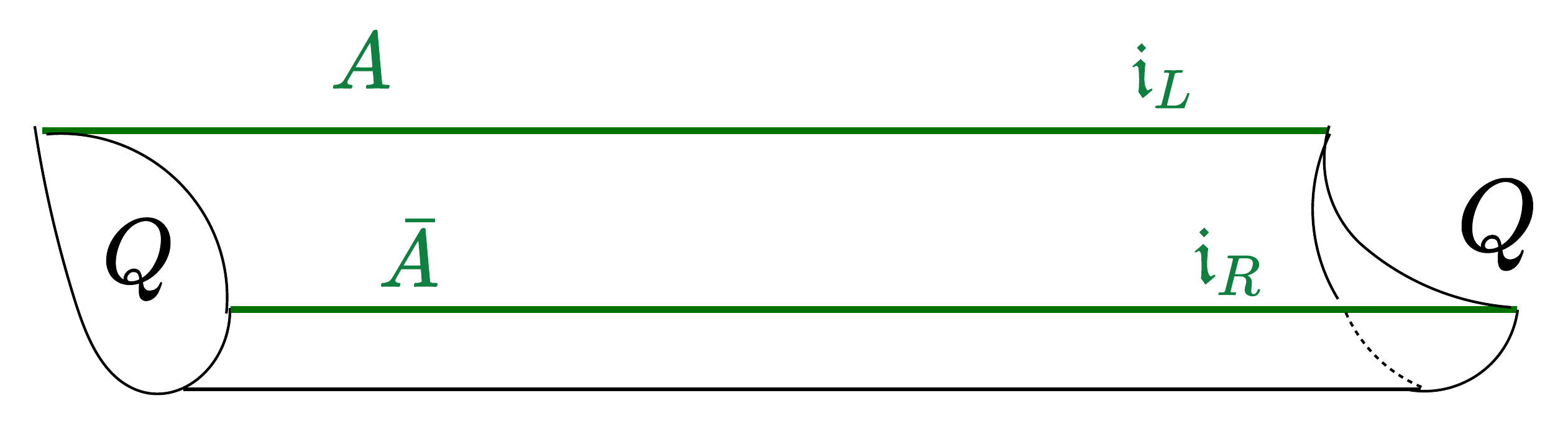}
    \caption{On the left, we show the bulk dual to the factorization map $i_{\text{boundary}}$  that acts on the AdS3 vacuum by introducing two ETW branes $Q$.  The disk like Cauchy slice $D$ is mapped to a strip $\tilde{D}$ with boundaries given by the two intervals and the ETW branes. On the right we applied the bulk diffeomorphism that corresponds to the conformal map to the boundary cylinder.   This shows how a two sided-BTZ black hole geometry arises the factorization of the AdS 3 vacuum  }
    \label{fig:planarbtz}
\end{figure}

 When the boundary manifold is a cylinder, the corresponding bulk gravity solution  has been determined (see e.g. \cite{Biswas:2022xfw}  \cite{Fujita:2011fp}).  The solution depends on the length $l$ of the cylinder, and experiences an analogue of the Hawking-page phase transition as $l$ increases past a critical value.   Our entanglement entropy problem corresponds to the long cylinder limit, where the results of \cite{Fujita:2011fp} imply that the bulk solution has the topology shown in the left of figure \ref{fig:bulkdouble}.   Cutting this geometry at the time-symmetric slice prepares the Hartle Hawking state $\ket{HH}$ on a bulk cauchy slice with the topology of a strip bounded by the Euclidean ETW branes (see right figure \ref{fig:planarbtz}).  
The Lorentzian evolution of this state produces a part of the planar two sided BTZ blackhole bounded by the Lorentzian evolution of the ETW branes \cite{May:2020tch}.   Heuristically, one can view the two sides of the black hole geometry as emerging from the two AdS Rindler Wedges of the original AdS3 vacuum.  The introduction of the ETW branes $Q$ is the bulk dual of the factorization map $i_{\text{boundary}}$,  which acts on the AdS3 vacuum as shown in the left of \ref{fig:planarbtz}.  


The gravitational Chern Simons formulation of AdS3/BCFT2 was introduced in \cite{Takayanagi:2020njm}, where the analogue of the boundary actions $I_{GH}$ and $I_{\text{brane}}$ was determined.  For RCFT's, the analogous bulk Chern Simons description of a the BCFT has also been worked out \cite{Fuchs:2001is}.  We appeal to these results as evidence that one can go beyond the classical solution described above to obtain a bulk TQFT path integral on the manifold $N \cup Q$ which reproduces the boundary character $\chi_{0}(-1/\tau)$. 
\subsection{Subregion states, bulk factorization map and the generalized entropy}
\label{sec:bulkfact}
The description of the bulk subregion Hilbert space and factorization map  is most straightforward in the unfolded picture, where $Z_{ee}(l)$ is interpreted as a solid torus path integral. 
As in our discussion of BH entropy, a bulk state counting interpretation of $Z_{ee}(l)$ is obtained by removing an infinitesmal solid tube and introduce a shrinkable boundary condition $e$ on the resulting boundary, which we identify as a bulk  entangling surface.  This gives a chiral theory on $\mathcal{A}\times S^1$ that is depicted in the left of figure \ref{fig:dfacthh}.   

This suggests we should define the bulk subregion Hilbert space just as in equation \eqref{sub}, using the representation matrix elements of $\SL_{q}^{+}(2,\mathbb{R})
$ as the wavefunctions : 
\begin{align}
    \mathcal{H}_{V}&= \int_{\oplus p \geq 0 } \,\,\dim_{q}p\,\, V_{p} \otimes \tilde{V}_{p} ,\nn
    V_{p}\otimes \tilde{V}_{p}&= \text{span} \{ \ket{p\, \mathfrak{i}_{L}\, s ; m^{L}_{k} }\} , \qquad    \braket{ g| p \, \mathfrak{i}_{L}\,s} = \sqrt{\dim_{q} p} \,\,R^{p}_{\mathfrak{i}_{L} \, s}(g) ,\qquad g\in \SL_{q}^{+}(2,\mathbb{R})
\end{align}
then the lower half of $\mathcal{A}\times S^1$ ( right of figure \ref{fig:dfacthh}) defines  defines the action of a factorization map $i_{\text{bulk}}$ that takes the Hartle Hawking state $\ket{HH}$ into a bulk extended Hilbert space $\mathcal{H}_{V}\otimes \mathcal{H}_{\bar{V}}$.   We can view the total bulk factorization map as a composition of $i_{\text{bulk}}$  and $i_{\text{boundary}}$ acting on the AdS vacuum

\begin{align}
     i_{\text{bulk}} \ket{HH} = i_{\text{bulk}} i_{\text{boundary}}\ket{\text{AdS vacuum}} =\int_{0}^{\infty} d p  \int_{-\infty}^{+\infty}ds \,     \sum_{m^{L}_{k}=m^{R}_{k}} e^{ \frac{-\pi^{2}}{l}(p^{2} + N_{m_{k}})}     \ket{p\, \mathfrak{i}_{L}s; m^{L}_{k}} \otimes \ket{p\, s\,\mathfrak{i}_{R}; m^{R}_{k}}.
\end{align}
By design, this factorization map gives the \emph{bulk} reduced density matrix
\begin{align}
   \rho_{V} = \int_{0}^{\infty} d p \int_{-\infty}^{+\infty}ds \, \sum_{m_{k}}   \ket{p\, \mathfrak{i}_{L}s; m_{k}} \otimes \bra{p\, s\,\mathfrak{i}_{L}; m_{k}}e^{-\frac{2\pi^{2}}{\ell} p^{2}}.
\end{align}
which satisfies $Z_{ee}(l)= \tr_{V} \rho_{V}$.

In the unfolded representation, we can once again appeal to the one loop exact nature of the bulk theory to conclude that the generalized entropy is given by the bulk entanglement entropy of $\rho_{V}$:
\begin{align}
S_{gen}= -\tr \rho_{V} \log \rho_{V} 
\end{align} 

\begin{figure}[h]
    \centering
    \includegraphics[width=0.75\textwidth]{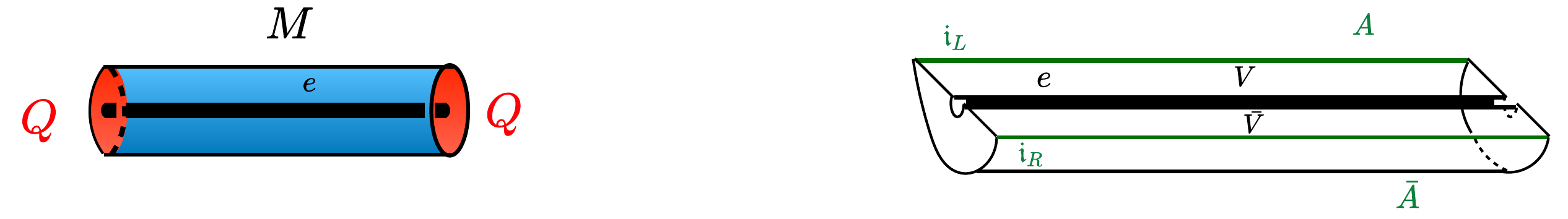}
    \caption{To obtain a bulk trace interpretation in the folded representation (left figure), we must excise a tube which ends on the ETQ branes denoted by $Q$.   The lower half of this geometry would then describe the factorization of the bulk Hartle Hawking state, as shown on the right.}
    \label{fig:facthh}
\end{figure}

What about the folded theory?  Here the bulk topology is more complicated. As shown in \ref{fig:facthh}, in order to separate the bulk Cauchy into $V$, and $\bar{V}$, the  bulk entangling surface carrying the quantum group edge modes would have to intersect the EOW branes Q.
It would be useful to work out the boundary terms in the action that would accommodate such an intersection.  We leave this for future work.

\subsection{Comparison with the c=1 Boson CFT}

In \cite{Hung:2019bnq}, the factorization problem was considered for the $c=1$ compact boson CFT.   To give some perspective to our calculation, It is useful to contrast the free boson factorization map with our holographic factorization defined above for $c>1$ CFT's.   We will contrast the nature of the edge modes in the two case.

First, note that in order to  leverage the extended chiral algebra of the compact boson, reference \cite{Hung:2019bnq} chose the  vacuum Ishibashi state \emph{with respect to the $U(1)$ Kacs-Moody  algebra} as the entanglement boundary state $\ket{e}$.  This satisfies the equation 
\begin{align}
(J_{n} - \bar{J}_{-n} )\ket{0}\rangle =0.
\end{align} 
 Propagation of this state in the closed string channel then produces the \emph{non degenerate} vacuum character of the $U(1)$ chiral algebra:
\begin{align}
    \langle \braket{0_{U(1)}| \tilde{q}^{L_{0} + \bar{L}_{0} -\frac{c}{12}}| 0_{U(1)} }\rangle =  \frac{\tilde{q}^{-(c-1)/24}}{\eta (q)} 
\end{align}
The reason we get the non degenerate character is that even though the null state $L_{-1}\ket{0}$ is still present, we also have the state $J_{-1} \ket{0}$.    The partition function in the open string channel is now obtained by applying the modular transform using the S-matrix of  $U(1)$ Kacs moody characters.  Equivalently, we  can expand $| 0_{U(1)} \rangle \rangle$ in terms of Cardy states which define local boundary conditions: these are given by  Neumann or Dirichlet boundary conditions.   Either way we obtain an open string partition function similar to \eqref{Zl} , but one which integrates over Dirichlet or Neumann boundary conditions with a \emph{constant} measure inherited from the modular S transform.   For example,  if we expand in terms of  Dirichlet cardy states,  one obtains an open string partition function
\begin{align}\label{DM}
        \mathtikz{ \pairA{0cm}{0cm} \copairA{0cm}{0cm} \draw (0cm,-0.1cm) node {\footnotesize $e$}; \draw (0cm,1cm) node {\footnotesize $e$}}= C    \int_{0}^{2 \pi R} d \phi_{a} \int_{ -\infty}^{\infty} d \phi_{b}  e^{\frac{-1}{l} (  \phi_{b}-\phi_{a }  )^{2}}  \eta^{-1}(q),
\end{align}
where $\phi_{a},\phi_{b}$ are the boundary values of the boson at the entangling surface and $R$ the boson radius.   These were refered to as CFT edge modes in  \cite{Hung:2019bnq}, since they parametrize different superselection sectors.  While these edge modes contribute a subleading term  $\log 2\pi R$  to the entanglement entropy, which is the analogue of the holographic edge mode term $\log S_{0}^P$ ,  the dominate part of the entanglement entropy comes from the descendants captured by $\eta(q)$. This is due to the well known fact that for $c\sim 1$, the Cardy density of states is dominated by the descendants, while for $c>>1$,  the Cardy density of states is dominated by primary states.

What if we choose the Virasoro Ishibashi state as the entanglement boundary state for the free boson?   Then we can repeat the same steps we followed for the holographic CFT, and obtain the same open string partition function as in \eqref{Zl}.  However, since the compact boson satisfies c=1, the entropy will be once again dominated by descendants and the density of states $\dim_{q}p$ would play a subleading role.  This makes sense, because the descendants are related to the Rindler oscillator modes which is responsible for the thermal entropy in the context of the Unruh effect for free bosons.

\section{Extended TQFT and 3d gravity}
\label{sec:ETQFT}
In this work, we defined subregion ``half wormhole" Hilbert space in terms of representations of a quantum group.  However, we have not given an explicit boundary condition that would lead to this Hilbert space via canonical quantization.  The reason is that, as explained  in \cite{Jafferis:2019wkd}, the shrinkable boundary condition in gravity must  be non local in the modular time variable.  This is because in contrast to QFT, in gravity the shrinkable boundary condition must implement the condition that the conical angle around the excised region is $2 \pi $.    This can be stated as a condition on the holonomy of the  spin connection around the excised region: 
\begin{align}
\oint  \omega = 2 \pi 
\end{align} 
which is manifestly a non local condition in modular time.   Since it is not clear how to quantize with such a non local boundary condition, we have formulated the subregion theory in a more abstract way: rather than specifying a condition on local fields, we specified edge modes transforming under a quantum group symmetry.  In our abstract approach, the validity of the shrinkable boundary condition is determined by the consistency of cutting and gluing  the gravitational path integral with these quantum group edge modes.

The mathematical framework that determines these sewing rules is given by  extended topological quantum field theory.  In this formalism, a boundary condition is specified by the choice of a boundary category.  This is analogous to how boundaries are specified in boundary conformal field theories, which can be defined abstractly without a Lagrangian. In this section we give an overview of this framework for the case of compact Chern Simons theory, and give a conjecture for its generalization to 3d gravity.  We interpret our computations of black hole entropy in section 2 and the QES formula in section 3 as one piece of evidence for this conjecture. 

\subsection{Extended TQFT}

   
   An extended TQFT in d spacetime dimensions is defined by a mapping $Z(\cdot )$ that assigns a mathematical object to surfaces of each codimension. In particular,  for surfaces of codimension zero, one and  two, the assignments are:
\begin{enumerate}
\item $Z(M_{d})$ = a partition function
\item  $Z(M_{d-1})$ = a complex vector space
\item $Z(M_{d-2})$ = a boundary ($\mathbb{C}$ linear) category 
\end{enumerate} 
Cobordisms- manifolds with initial and final boundaries- correspond to maps between the structures assigned to these boundaries.  For cobordisms between $d-1$ manifolds $X$ and $Y$, these are linear maps $Z(X) \to Z(Y)$ describing a quantum evolution between initial and final vector spaces. Similarly a cobordism between d-2 manifoilds $W$ and $V$ is assigned to a functor, i.e. a map between categories $Z(W)$ and $Z(V)$.  This rule also applies to cobordism into the empty manifold:  A cobordism from the empty $d-1$ manifold to $M_{d-1}$ is a linear  map $
\mathbb{C} \to Z(M_{d-1})$, which is a choice of a vector $Z(M_{d-1})$ :in the usual language this is the preparation of a state by the path integral.

Notice that as we move up in co-dimension, the mathematical structure becomes more refined and contains more information about the theory.   In fact, given the assignments at codimension $k$, one can reconstruct the assignments at co dimension $n<k$ and therefore reproduce the whole theory. 
\begin{figure}[h]
    \centering
    \includegraphics[width=0.75\textwidth]{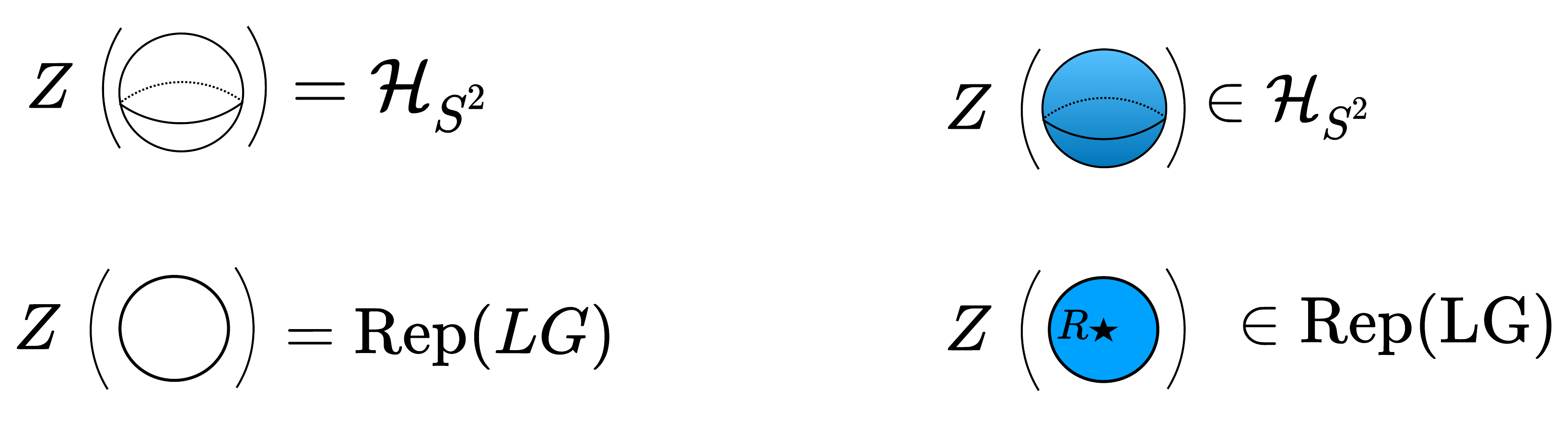}
    \caption{Some of the path integral assignments for Chern Simons theory with compact gauge group. }
    \label{fig:CStqft}
\end{figure}
For the purposes of computing entanglement measures, it suffice to extend down to tier 2.
In this formalism, the boundary category $Z(M_{d-2})$ determines the edge modes that live on a spatial boundary, and replaces the conventional notion of a boundary condition.  
\subsection{Compact Chern Simons}
Let us illustrate this reconstruction process in d=3 by considering Chern Simons theory with compact gauge group $G$.   In particular, we will explain how the abstract boundary condition reproduces the Hilbert spaces obtained by quantizing 
with the WZW model boundary condition 
\begin{align}
    A_{0}=0,
\end{align} with 0 labeling the time coordinate. In extended TQFT,  this boundary condition is given by 
\begin{align} 
Z(M_{d-2})= Z(S^1)=\text{Rep(LG)},
\end{align}
where Rep(LG) is the representation category of the loop group LG.  Figure \ref{fig:CStqft} shows the some of the assignments that result from this choice.   Recall the connection between the different tiers.  Once we have chosen $Z(S^2)=\mathcal{H}_{S^2}$ to be the sphere Hilbert space, a ball $B^3$ bounded  by $S^2$ must be assigned to an element $\ket{Z(B^3)}\in Z(S^2)$.
A similar relationship appears at one level higher in codimension:  Since $Z(S^1)=$ Rep(LG), a spatial manifold with punctures bounded by $S^1$  is assigned to an element of Rep(LG) ( the puncture determines which element) .  The latter is just the Hilbert space of a disk, which consists of WZW model edge modes on the circular boundary.

\begin{figure}[h]
    \centering
    \includegraphics[width=0.75\textwidth]{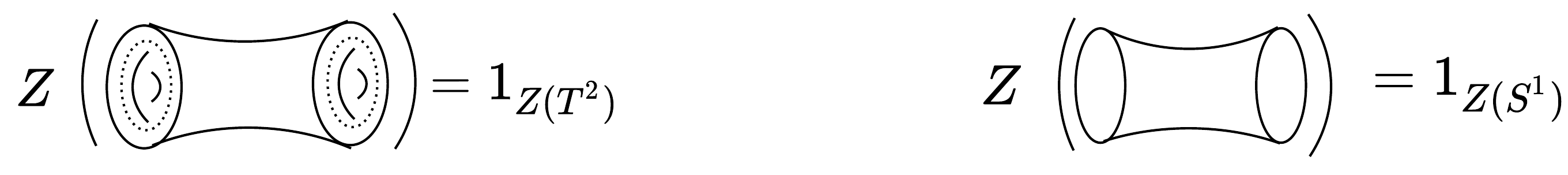}
    \caption{A Spacetime wormhole (left) is mapped to a WZW partition function, where as a spatial wormhole (right) is mapped to a Hilbert space.   }
    \label{wh}
\end{figure}

Next consider the wormhole geometries in figure 
\ref{wh}.
  The TQFT rules dictate that a $T^{2} 
\times I$ cobordism is assigned to the evolution operator $e^{-tH}$ on $Z(T^2)$: since $H=0$ in a TQFT this is the identity operator.    Since $Z(T^{2})$ is the space of WZW character $\ket{\chi_{p_{i}}}$ , the resolution of identity 
\begin{align}\label{one}
    \mathbb{1} = \sum_{i} \ket{\chi_{p_{i}}} \bra{\chi_{p_{i}}}     
\end{align}
corresponds to the partition function of the WZW model on the torus. 
Similarly, the extended TQFT assigns to an $S^{1} 
\times I$ cobordism the identity functor on Rep(LG), generalizing \eqref{one}.    Just as the identity linear map has a resolution of of identity $1= \sum_{R} \ket{R}\bra{R}$ over a basis $\ket{R}$, the identity functor has the ``resolution of of identity" given by the tensor product \emph{Hilbert space} 
\begin{align}\label{1F}
 \mathbb{1}_{\text{Rep(LG)}}= \oplus_{R} V_{R}\otimes V_{R}^*
\end{align} 
where $V_{R}$ are integrable representations of LG\footnote{We thank Ahsan Khan for explaining this point to us.}.   Once again, this Hilbert space agrees with the conventional one obtained via canonical quantization with the WZW boundary condition \cite{Elitzur:1986ye}.  
\paragraph{Cutting and gluing Hilbert spaces}
In ordinary TQFT, a partition function $Z(M)$ on a closed d manifold $M= \bar{M}_{L}\cup M_{R}$  can be computed by cutting along the  $S^1$ boundaries  separating  $M_{L}$ and $M_{R}$, and then computing the inner product\footnote{The most basic axioms of a TQFT doesn't include an inner product, in which case $(,) $ is just the natural pairing of a vector and its dual.   In physics we usually assume an inner product structure which we can use for the gluing }
\begin{align}
    Z(M) = (Z(M_{L}),Z(M_{R}) )
\end{align}
 on the Hilbert space of the  $S^1$ boundaries.  To describe the factorization of Hilbert spaces, we apply the same principle in one co-dimension higher.  e.g. for Chern Simons theory we can compute the Hilbert space   $Z(S^2)$ by using an ``inner product"  (,) on Rep(LG) \footnote{See \cite{freed1993extended} for a definition of this inner product on Rep(G), with G a finite group. The categorical definition of this inner product on a pair of representations $X \in$ Rep(LG)  and $Y\in$  Rep(LG)$^{op}$is the set of homomorphisms 
\begin{align}
    (X,Y)=\text{Hom}_{\text{Rep(LG)}} (X,Y)
\end{align}
}.  This is defined as a fusion of a representation and its conjugate by projecting onto the invariant subspace \cite{Wong:2017pdm}.: 
 \begin{align}\label{singlet}
 Z(S^{2}) &=  ( Z(D^{2}),Z(D^{2} ))\\
          &\simeq  Z(D^{2}) \otimes_{LG} Z(D^{2}) .  
 \end{align}   
In the second line, we have expressed this fusion as a relative tensor product, with the $\otimes_{LG}$ denoting a quotient that projects onto the singlets under $LG$ acting simultaneously on the boundaries of the two disks (this is the Gauss law constraint that crosses the entangling surfaces).   This reproduces the standard one dimensional Hilbert space $Z(S^{2})$.   One can repeat the same computation for the torus Hilbert space by making two cuts:
\begin{align}
Z(T^{2}) =   \left(Z(S^{1})\times I \right) \otimes_{LG\otimes LG} \left(Z(S^{1})\times I \right).
\end{align} 
Here the gluing of the wormhole Hilbert spaces \eqref{1F} gives one singlet state per representation $R$. 

\paragraph{Shrinkability of the WZW boundary condition}
The discussion above shows that the Rep(LG) defines a good entanglement boundary condition because it leads to WZW edge modes that consistently fuse together Hilbert spaces along circles.  This naturally leads us to define Hilbert space factorization as an inverse of the fusion product: starting with the physical Hilbert space, identified with a relative tensor product as in\eqref{singlet}, we factorize it by lifting the projection onto the singlet states.  However, this does not lead to well defined reduced density matrices.  To see the problem, consider the factorization of $Z(S^2)$. Explicitly lifting the singlet projection in \eqref{singlet} leads to an Ishibashi state:
\begin{align} \label{ishi}
    i:Z(S^{2}) &\hookrightarrow Z(D^{2}) \times Z(D^{2}) \nn
    \ket{Z(B^3)} &\to
        \sum_{m} \ket{ m}\otimes \ket{ \bar{m}} ,
\end{align}
where $m$ labels the Kacs Moody descendants of the identity.
However the non-normalizability of the Ishibashi state prevents us from obtaining a normalizable density matrix when we trace over one side. 

To deal with this issue, we introduce a new ingredient into the extended TQFT formalism that we referred to as the shrinkable boundary condition.    To regularize the UV divergences, we introduce a stretched entangling surface that separates the two disks that glue into $S^2$ and define a regulated mapping $i_{\epsilon}$.  
The geometry that realizes  the state $i_{\epsilon}\ket{Z(B^3)}$ is now given by a solid ball with a neighborhood of the entangling surface removed.   This is a cobordism between the empty manifold and two disks, which is a choice of a state
\begin{align}
i_{\epsilon}\ket{Z(B^3)} \in Z(D^{2}) \times Z(D^{2})
\end{align}
 By the state channel duality (identifying one copy of $Z(D^{2})$ with its dual$Z(D^{2})^*$ ) , this geometry can also be viewed as cobordism from a disk to an oppositely oriented one, so it is assigned to the WZW evolution operator on a circle of size $l$: 
\begin{align}
    i_{\epsilon}\ket{Z(B^3)} \simeq  e^{\frac{- \pi^2 \epsilon}{l} (L_{0}+\bar{L_{0}} -\frac{c}{12})} \in Z(D^{2}) \times Z(D^{2})^*
\end{align}
The corresponding factorized state is given by
\begin{align}
   i_{\epsilon}\ket{Z(B^3)}  =\frac{1}{\sqrt{Z( \epsilon)}}
        \sum_{m} e^{\frac{- \pi^2 \epsilon}{l} (L_{0}+\bar{L_{0}})} \ket{ m}_{V} \otimes \ket{ \bar{m}}_{\bar{V}} ,
\end{align}
where we have included a normalization factor that leads to a normalized density matrix.  

In this context, the shrinkability is a consistency check that our regularization procedure has not alterred the correlations of the original, unfactorized state.  This requires that 
\begin{align}
  Z(S^{3}) &= (Z(B^3), Z(B^3)) \nn
&\simeq \lim_{\epsilon \to 0}  \bra{Z(B^3)} i^{\dagger}_{\epsilon} i_{\epsilon}\ket{Z(B^3)} ,
\end{align}
Here the adjoint map $i^{\dagger}_{\epsilon} $ is a regulated version of the projection into singlets , and the symbol $\simeq$ to denote equality up to a counterterm subtraction that does not change the entropy.

We interpret this equation as a consistency constraint that ensures the same $Z(S^3)$ can be computed either two different ways.  The first way computes $Z(S^3)$ as the norm of a state by gluing the 3 ball to itself to make a 3 sphere.   The second way computes $Z(S^3)$ as a trace
\begin{align}
    Z(S^{3})=\lim_{\epsilon \to 0}  \bra{Z(B^3)} i^{\dagger}_{\epsilon} i_{\epsilon}\ket{Z(B^3)} = \lim_{\epsilon \to 0} \tr_{Z(D^2)} e^{\frac{- \pi^2 \epsilon}{l} (L_{0}+\bar{L_{0}}-c/12)} ,
\end{align}
since the regulated geometry is  $S_{3}$ minus a solid torus, which has a non contractible Euclidean time circle. 

\paragraph{Other boundary conditions}
Finally, we note we can incorpporate different types of boundary condition into extended TQFT, and not all of them have to be shrinkable.   For example, Chern Simons theory has  gapped, topological boundary conditions that do not lead to gapless edge modes, and therefore cannot be glued to other gapped boundaries.  To fully describe the theory,  we introduce different types of  co-dimension 2 surfaces with a boundary label to distinguish the different boundary conditions.      Once we allow different boundary conditions, there can also be interfaces between different boundaries.  These additions to the extended TQFT structure will be relevant to our 3d gravity applications. 

\subsection{3d gravity (on  a fixed topology) as an extended TQFT}

Our approach to entanglement entropy in 3d gravity is based on the proposal that the bulk theory can be formally defined as an extended TQFT, following the framework outlined above.  In fact, recent work has given strong evidence that 3d gravity amplitudes can be computed via Virasoro TQFT \cite{Collier:2023fwi}\footnote{This is just a new name for Teichmuller TQFT } .   This was formulated as an un-extended TQFT  which assigns a Hilbert space of Liouville conformal blocks to co-dimension 1 surfaces.    Our work pushes this further by proposing an extension of the Virasoro TQFT to co-dimension2 that is consistent with gravitational entropies.

In our application of extended TQFT to 3d gravity, we 
 introduced two types of boundary circles, one at infinity assigned to the category Rep(Vir) of Virasoro representations, and one in the bulk assigned to the category  Rep $\SL_{q}^{+}(2,\mathbb{R})$ of quantum group representations.   The led to two different types of spatial wormholes and associated Hilbert spaces.   
\begin{figure}[h]
    \centering
    \includegraphics[width=0.75\textwidth]{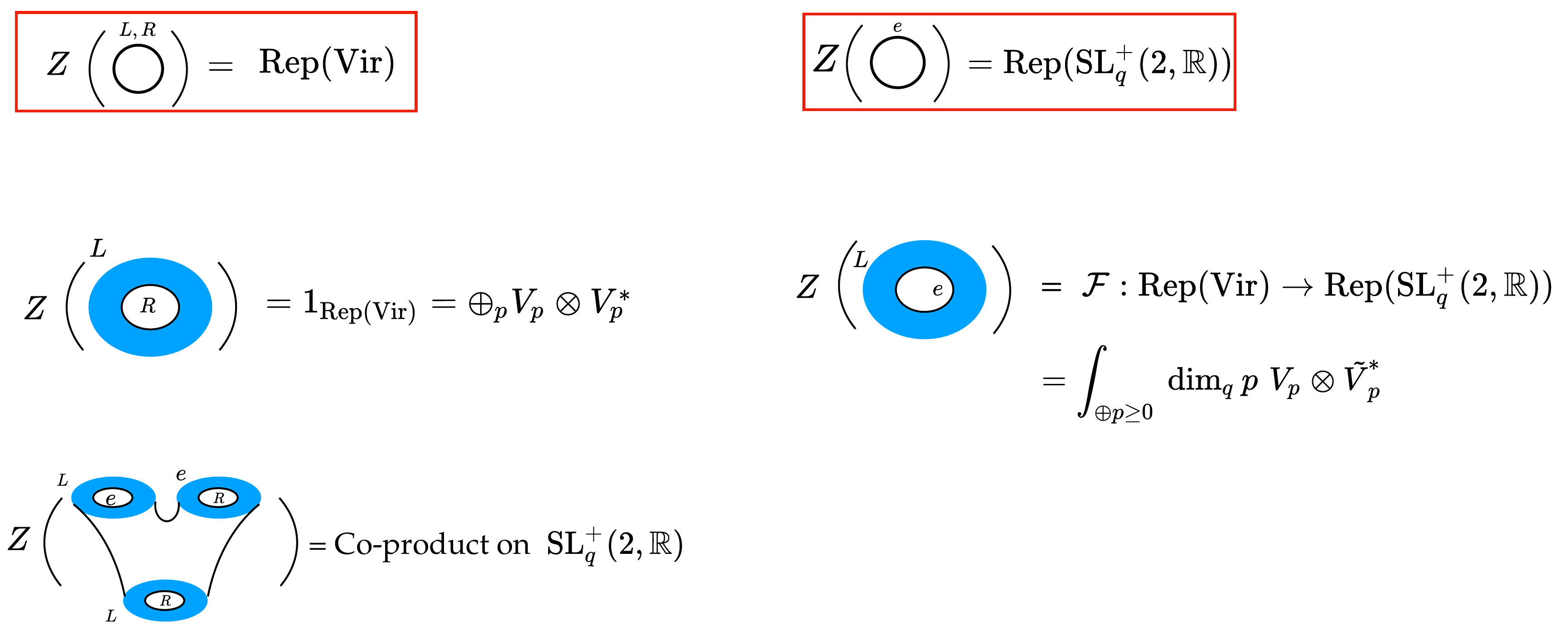}
    \caption{Path integral assignments for the gravitational  extended TQFT.}
    \label{fig:gravtqft}
\end{figure}
First, the two sided Hilbert space \eqref{H} on the ER bridge is given by identity functor on Rep(Vir):
\begin{align}
\mathbb{1}_{\text{Rep(Vir)}} = \oplus_{p} V_{p}\otimes V_{p}^*
\end{align} 
This is the direct analogue of \eqref{1F}.
Similarly, the ``half wormhole" Hilbert space is the functor  $\mathcal{F}$ from Rep(Vir) into Rep( $\SL_{q}^{+}(2,\mathbb{R})$) discovered by Teschner as a solution of the Liouville modular bootstrap:
\begin{align}
\mathcal{F} &: \text{Rep(Vir)} \to \text{Rep} ( \SL_{q}^{+}(2,\mathbb{R}))\nn
\mathcal{F} &=\int_{\oplus p \geq 0} \,\dim_{q}p   \, V_{p} \otimes \tilde{V}_{p} 
\end{align} 

Generalizing the extended TQFT framework for compact Chern Simons theory, we define the fusion of two half wormhole Hilbert spaces via a relative tensor product over the quantum group $\SL_{q}^{+}(2,\mathbb{R}))$
Thus , the bulk factorization map \eqref{ibulk}, should be interpreted in relation to the Hilbert space fusion 
\begin{align}
    \mathcal{H} = \mathcal{H}_{V}\otimes_{\SL_{q}^{+}(2,\mathbb{R}))} \mathcal{H}_{\bar{V}},
\end{align}
in direct analogy with \eqref{singlet} and our discussion below \eqref{ishi}.  

For our proposed extended TQFT to be consistent, the same boundary condition must be used to factorize  along different types of entangling surfaces, and in different states.   We view 
our computation of BH entropy and the QES formula as evidence for this consistency.  However, we should note that the computation of the QES formula made use of an interface between a shrinkable boundary and an EOW brane.  While boundary interfaces are a feature of TQFT's , a proper Virasoro TQFT formulation of the EOW brane is still lacking.

\section{Conclusions}
We have shown that the QES formula in pure AdS3 gravity can be interpreted as bulk entanglement entropy.  The leading area term describes the entanglement entropy of gravitational edge modes that transform under $\SL_{q}^{+}(2,\mathbb{R})$.   These edge modes are associated to bulk subregion wavefunctions given by representation matrices of $\SL_{q}^{+}(2,\mathbb{R})$, and the  bulk factorization map is identified with the co-product on $\SL_{q}^{+}(2,\mathbb{R})$. 
This supports the perspective that AdS3 gravity can be viewed as a topological phase in which the BH entropy is identified with the entanglement entropy of gravitatonal anyons.

The fact that the same $\SL_{q}^{+}(2,\mathbb{R})$ symmetry also governs the edge modes and BH entropy of the two sided BTZ black hole suggests the existence of a bulk path integral equipped with a consistent notion of   cutting and gluing along co-dimension 2 surfaces.   As advocated in \cite{Mertens:2022ujr}  (see also \cite{Donnelly:2018ppr}),  extended TQFT provides the appropriate categorical framework to describe such a path integral.  This formulation of the bulk theory provides a sharp answer to the question posed in the introduction: why does the Euclidean path integral give the correct state counting?   In this formulation, the bulk  $\SL_{q}^{+}(2,\mathbb{R})$ edge modes are part of the data which \emph{defines} the path integral.    The shrinkable boundary condition is a sewing relation that ensures that the Gibbons-Hawking calculation, defined in the ``closed sector" of the TQFT (with no bulk boundaries),  must be consistent with a trace calculation in the open sector of the TQFT.     The area term of the QES formula is then simply the Plancherel measure which forms  part of the data of Rep($\SL_{q}^{+}(2,\mathbb{R})$).
 
\paragraph{Future directions}
Much work remains to be done in establishing the validity of our proposal for the bulk extended TQFT.   First we should compute entanglement entropy in more general states created with operator insertions and with more general subregions.    In these cases, the entanglement entropy is no longer determined by conformal symmetry alone.   However, the emergence of bulk gravity can still be attributed to the dominance of the vacuum block \cite{Asplund:2014coa} \cite{Asplund:2015eha} .   This suggests that our bulk TQFT remains valid, although we may have to introduce Wilson lines to account for the operator insertions. 

One drawback of our abstract description is that it does not give  an explicit quantization scheme that produces the quantum group representations.  In the classical setting, there is a hint of how this might be achieved.    Using the first order (BF) formulation of gravity, \cite{Dupuis:2020ndx} succeeded in obtaining  \emph{classical} quantum group edge modes by deforming the gravity action  by a boundary term.  Unfortunately, their quantum group was $\mathcal{U}_{q}(SL(2))$ rather than $\SL^+_{q}(2,\mathbb{R})$ \footnote{$\SL^+_{q}(2,\mathbb{R})$ is related to $\mathcal{U}_{q}(SL(2))$ by the modular double construction \cite{Faddeev:1999fe}.}.  Nevertheless, understanding the relation of  our quantum group edge modes to to those of \cite{Dupuis:2020ndx} might shed light on the proposal that gravity edge modes arise from large diffeomorphisms of a subregion. 
 
 More recently, in compact Chern Simons theory,  \cite{Ronak} have applied an explicit (spatially) non local boundary condition to obtain a subregion density matrix.  The resulting entanglement  entropy consist of the log of the quantum dimension of Wilson lines passing through the entangling surface.  This suggests an interpretation as counting quantum group edge modes, providing a compact gauge theory analogue of our 3d gravity construction. 

Going beyond our proposal for AdS3 gravity, it is natural to ask whether a similar approach can provide a canonical interpretation of de Sitter entropy.   Another natural generalization is to consider charged black holes, which corresponds to having an extended chiral algebra in the boundary theory.     Given that both the structure of extended TQFT and our argument for ``bulk modular invariance" carries over in arbitrary dimensions, we should also consider how these structures can be leveraged to give a better understanding of the Gibbons Hawking calculation in higher dimensions.  

Finally, the ultimate question we would like to answer is how to define entanglement entropy directly in the bulk string theory.  Some progress was made along these lines in the context of the A model  topological string \cite{Donnelly:2020teo, Jiang:2020cqo}.   Here, bulk modular invariance and the extended TQFT framework can be applied directly to the closed string partition function on the resolved conifold, which is a fiber bundle over a sphere.  In this case, the ``local holography" which replaces a small disk on the sphere with the bulk edge modes takes the form of an open-closed string duality (related but not exactly the same as the Gopakumar -Vafa duality).  Perhaps not surprisingly, the bulk edge modes correspond to the D branes of the topological string, with a special value of the worldvolume holonomy needed to satisfy the shrinkable boundary condition.  It would be interesting to apply these ideas to a physical string theory, and find an explicit realization of these ``entanglement branes".

\section*{Acknowledgements}
It is a pleasure to thank  Hao Geng, Daniel Jafferis, Ahsan Khan, David Kolchmeyer, Samir Mathur, Thomas Mertens, Joan Simon, and Ronak Soni  for discussions related to this paper.   
GW is supported by the Harvard Center of Mathematical Sciences and Applications at Harvard University, and thanks the Aspen Center for Physics for hospitality.   
\appendix
\section{Appendix A: A review of the Peter Weyl Theorem}
\subsection{ SL$^+_{q}(2,\mathbb{R})$ and its representations}
 The quantum semi-group SL$^+_{q}(2,\mathbb{R})$ plays a key role in the bulk derivation of BH entropy and the RT formula in 3d gravity.   It is best defined via its representation category.   Below we explain this point of view using the Peter Weyl theorem.
\paragraph{The Peter Weyl theorem} 
 In the usual formulation of a symmetry, we begin with a group $G$ satisfying the group axioms, and then define representations of $G$.   The representations of $G$ form a category $\text{Rep}(G)$, defined by a set of data which includes the irreducible representations, their fusion rules, and their dimensions. It turns out one can reconstruct the symmetry $G$ from the data which defines its representation category.  This abstract point of view provides a useful way to define the quantum semi-group $SL^+_{q}(2,\mathbb{R})$.  

To see how this works, lets start with the case of an ordinary compact Lie group $G$.  Properties of $G$ can be characterized by the space $L^{2}(G)$ of square integrable functions on $G$. $L^{2}(G)$ is a Hopf algebra under pointwise multiplication, and properties of this algebra encode the group axioms satisfied by $G$.   The Peter Weyl theorem gives a characeterization of this algebra in terms of Rep $G$ data:
\begin{align}\label{PeterW}
    L^{2}(G) =\oplus_R V_{R}\otimes V_{R^*}
\end{align}
where $V_{R}$ are the irreps of $G$ and $V_{R}^*$ are the conjugate representations.  This equation arises from the decomposition of the  regular representation of the group $G$ on $L^{2}(G)$, which acts by left and right multiplication on the argument of a function $f \in L^{2}(G)$ 
\begin{align}\label{action}
    f(g) \to f(h_{L}g h_{R}^{-1}) ,\quad  h_{L},h_{R} \in G
\end{align}
Concretely, equation \ref{PeterW}  means that a basis for $L^{2}(G)$ is given by the representation matrix elements $$R_{ab}(g),\qquad g\in G,\qquad a,b=1,\cdots \dim R ,$$ since $a$ and $b$ transform under $V_{R}$ and $V_{R}^{*}$ respectively.  The RHS of \eqref{PeterW} determines the Hopf algebra structure of of $L^{2}(G)$ via data associated to Rep (G): for example, pointwise multiplication of the Hopf algebra $L^{2}(G)$ is captured by the Clesbch Gordon coefficient of Rep $G$ which determine the decomposition of $R_{ab}(g) \cdot R'_{cd}(g')$ into irreps.  Note that the representations $V_{R}$ can also be viewed as representations of $\mathcal{U}(G)$, the universal enveloping algebra generated by the Lie algebra elements of $G$, with a multiplication rule that is consistent with the Lie algebra commutator.

For ordinary groups, the Peter Weyl theorem admits two generalizations. First when $G$ is non compact, the RHS  in general includes a continuous set of representations with an appropriate Plancherel measure $d \mu (R)\equiv \dim R$.  This measure appears in the orthogonality relation for the matrix elements, defined with respect to the haar measure: 
\begin{align}
    \int dg R_{ab}(g) R^{*}{cd}(g) =\frac{\delta_{ac} \delta_{bd}}{\dim R}
\end{align}

Then $L^{2}(G)$ is given by
\begin{align}
      L^{2}(G) =\int_{\oplus R}  d \mu(R)  V_{R}\otimes V_{R^*}
\end{align}
Second, give a subgroup $H$, the Peter Weyl theorem can be applied to the coset $G/H$.  In this case the we have
\begin{align}\label{coset} 
    L^{2}(G/H)= \int_{\oplus R}d\mu(R) V_{R} \otimes V_{R,0}^*,
\end{align}
where $V_{R,0}^*$ denotes a projection on to a subspace invariant under the right multiplication action of $H$.  In 3d gravity, the bulk subregion states the form of the RHS of \eqref{coset}.
\paragraph{The representation category for SL$^+_{q}(2,\mathbb{R})$}

The Peter Weyl theorem and its generalizations \eqref{coset}, suggests we can define more generalized forms of symmetries \footnote{e.g. these symmetries need not be invertible} by starting with the representation data on the RHS, and then finding the $L^{2}$ space for which these representation forms a complete set.  This is in essence how Teschner arrived at his definition of the quantum semi group $SL^+_{q}(2,\mathbb{R})$.  Specifically, he considered the representations of the modular double of the q-deformed universal enveloping algebra $\mathcal{U}_{q}(\SL(2))$, with $q=e^{i \pi b^{2}}, b\in \mathbb{R}$.  They form a continuous set of representations which we denote by $V_{p}$, with $p\in \mathbb{R}$. These are representations which appeared in the solution to the modular bootstrap for Liouville theory.  They satisfy special properties because they are simultaneously representations of $\mathcal{U}_{q}(\SL(2))$ with $\tilde{q}=e^{i \pi/b^{2}}$.
The Plancherel measure associated to these representations is
\begin{align}
    d \mu(p) \equiv \dim_{q}p = \sqrt{2} \sinh ( 2 \pi b p ) \sinh ( 2 \pi b^{-1} p )
\end{align}
We can then define SL$^{+}_{q}(2,\mathbb{R})$ as the space on which the Hopf Algebra $L^{2}(SL^{+}_{q}(2,\mathbb{R}))$ of square integrable functions has the spectral decomposition:
\begin{align}
    L^{2}(SL^{+}_{q}(2,\mathbb{R})) =\int_{\oplus p\in \mathbb{R}^+}  d \mu(p)  V_{p}\otimes V_{p^*}
\end{align}
Ip has proved this generalized Peter Weyl theorem, and in particular provided explicit formulas for the representation basis elements implied by the right hand side.  These matrix elements have continuous indices $s \in \mathbb{R}$ and satisfy the usual representation property 
\begin{align}
    R^{p}_{ab}(g_{1}g_{2})=\int_{-\infty}^{\infty}  d s \,  R^{p}_{as}(g_{1})R^{p}_{sb}(g_{2}) 
\end{align}

\newpage

\bibliographystyle{JHEP-2}
\bibliography{3Dgrav}
\end{document}